
%
%
%
\def\unredoffs{} \def\redoffs{\voffset=-.31truein\hoffset=-.59truein}
\def\speclscape{\special{ps: landscape}}
%
%
%
%
\newbox\leftpage \newdimen\fullhsize \newdimen\hstitle \newdimen\hsbody
\tolerance=1000\hfuzz=2pt
\catcode`\@=11 
\def\bigans{b }
\def\answ{b }
\ifx\answ\bigans\message{}  
\magnification=1000\unredoffs\baselineskip=16pt plus 2pt minus 1pt
\hsbody=\hsize \hstitle=\hsize 
\else\message{(This will be reduced.} \let\l@r=L
\magnification=1000\baselineskip=16pt plus 2pt minus 1pt \vsize=7truein
\redoffs \hstitle=8truein\hsbody=4.75truein\fullhsize=10truein\hsize=\hsbody
\output={\ifnum\pageno=0 
  \shipout\vbox{\speclscape{\hsize\fullhsize\makeheadline}
    \hbox to \fullhsize{\hfill\pagebody\hfill}}\advancepageno
  \else
  \almostshipout{\leftline{\vbox{\pagebody\makefootline}}}\advancepageno
  \fi}
\def\almostshipout#1{\if L\l@r \count1=1 \message{[\the\count0.\the\count1]}
      \global\setbox\leftpage=#1 \global\let\l@r=R
 \else \count1=2
  \shipout\vbox{\speclscape{\hsize\fullhsize\makeheadline}
      \hbox to\fullhsize{\box\leftpage\hfil#1}}  \global\let\l@r=L\fi}
\fi
%
\newcount\yearltd\yearltd=\year\advance\yearltd by -1900

\def\Title#1#2{\nopagenumbers\abstractfont\hsize=\hstitle\rightline{#1}%
\vskip 1in\centerline{\titlefont #2}\abstractfont\vskip .5in\pageno=0}
\def\Date#1{\vfill\leftline{#1}\tenpoint\supereject\global\hsize=\hsbody%
\footline={\hss\tenrm\folio\hss}}
%

\def\draftmode{\message{ DRAFTMODE }\def\draftdate{{\rm preliminary draft:
\number\month/\number\day/\number\yearltd\ \ \hourmin}}%
\headline={\hfil\draftdate}\writelabels\baselineskip=20pt plus 2pt minus 2pt
 {\count255=\time\divide\count255 by 60 \xdef\hourmin{\number\count255}
  \multiply\count255 by-60\advance\count255 by\time
  \xdef\hourmin{\hourmin:\ifnum\count255<10 0\fi\the\count255}}}
\def\nolabels{\def\wrlabeL##1{}\def\eqlabeL##1{}\def\reflabeL##1{}}
\def\writelabels{\def\wrlabeL##1{\leavevmode\vadjust{\rlap{\smash%
{\line{{\escapechar=` \hfill\rlap{\sevenrm\hskip.03in\string##1}}}}}}}%
\def\eqlabeL##1{{\escapechar-1\rlap{\sevenrm\hskip.05in\string##1}}}%
\def\reflabeL##1{\noexpand\llap{\noexpand\sevenrm\string\string\string##1}}}
\nolabels
%
\global\newcount\secno \global\secno=0
\global\newcount\meqno \global\meqno=1
\def\newsec#1{\global\advance\secno by1\message{(\the\secno. #1)}
\global\subsecno=0\eqnres@t\noindent{\bf\the\secno. #1}
\writetoca{{\secsym} {#1}}\par\nobreak\medskip\nobreak}
\def\eqnres@t{\xdef\secsym{\the\secno.}\global\meqno=1\bigbreak\bigskip}
\def\sequentialequations{\def\eqnres@t{\bigbreak}}\xdef\secsym{}
\global\newcount\subsecno \global\subsecno=0
\def\subsec#1{\global\advance\subsecno by1\message{(\secsym\the\subsecno. #1)}
\ifnum\lastpenalty>9000\else\bigbreak\fi
\noindent{\it\secsym\the\subsecno. #1}\writetoca{\string\quad
{\secsym\the\subsecno.} {#1}}\par\nobreak\medskip\nobreak}
\def\appendix#1#2{\global\meqno=1\global\subsecno=0\xdef\secsym{\hbox{#1.}}
\bigbreak\bigskip\noindent{\bf Appendix #1. #2}\message{(#1. #2)}
\writetoca{Appendix {#1.} {#2}}\par\nobreak\medskip\nobreak}
%
%
\def\eqnn#1{\xdef #1{(\secsym\the\meqno)}\writedef{#1\leftbracket#1}%
\global\advance\meqno by1\wrlabeL#1}
\def\eqna#1{\xdef #1##1{\hbox{$(\secsym\the\meqno##1)$}}
\writedef{#1\numbersign1\leftbracket#1{\numbersign1}}%
\global\advance\meqno by1\wrlabeL{#1$\{\}$}}
\def\eqn#1#2{\xdef #1{(\secsym\the\meqno)}\writedef{#1\leftbracket#1}%
\global\advance\meqno by1$$#2\eqno#1\eqlabeL#1$$}
%
\newskip\footskip\footskip14pt plus 1pt minus 1pt 
\def\footnotefont{\ninepoint}\def\f@t#1{\footnotefont #1\@foot}
\def\f@@t{\baselineskip\footskip\bgroup\footnotefont\aftergroup\@foot\let\next}
\setbox\strutbox=\hbox{\vrule height9.5pt depth4.5pt width0pt}
\global\newcount\ftno \global\ftno=0
\def\foot{\global\advance\ftno by1\footnote{$^{\the\ftno}$}}
%
\newwrite\ftfile
\def\footend{\def\foot{\global\advance\ftno by1\chardef\wfile=\ftfile
$^{\the\ftno}$\ifnum\ftno=1\immediate\openout\ftfile=foots.tmp\fi%
\immediate\write\ftfile{\noexpand\smallskip%
\noexpand\item{f\the\ftno:\ }\pctsign}\findarg}%
\def\footatend{\vfill\eject\immediate\closeout\ftfile{\parindent=20pt
\centerline{\bf Footnotes}\nobreak\bigskip\input foots.tmp }}}
\def\footatend{}
%
%
\global\newcount\refno \global\refno=1
\newwrite\rfile
\def\ref{[\the\refno]\nref}
\def\nref#1{\xdef#1{[\the\refno]}\writedef{#1\leftbracket#1}%
\ifnum\refno=1\immediate\openout\rfile=refs.tmp\fi
\global\advance\refno by1\chardef\wfile=\rfile\immediate
\write\rfile{\noexpand\item{#1\ }\reflabeL{#1\hskip.31in}\pctsign}\findarg}
\def\findarg#1#{\begingroup\obeylines\newlinechar=`\^^M\pass@rg}
{\obeylines\gdef\pass@rg#1{\writ@line\relax #1^^M\hbox{}^^M}%
\gdef\writ@line#1^^M{\expandafter\toks0\expandafter{\striprel@x #1}%
\edef\next{\the\toks0}\ifx\next\em@rk\let\next=\endgroup\else\ifx\next\empty%
\else\immediate\write\wfile{\the\toks0}\fi\let\next=\writ@line\fi\next\relax}}
\def\striprel@x#1{} \def\em@rk{\hbox{}}
\def\lref{\begingroup\obeylines\lr@f}
\def\lr@f#1#2{\gdef#1{\ref#1{#2}}\endgroup\unskip}
\def\semi{;\hfil\break}
\def\addref#1{\immediate\write\rfile{\noexpand\item{}#1}} 
\def\startrefs#1{\immediate\openout\rfile=refs.tmp\refno=#1}
\def\xref{\expandafter\xr@f}\def\xr@f[#1]{#1}
\def\refs#1{\count255=1[\r@fs #1{\hbox{}}]}
\def\r@fs#1{\ifx\und@fined#1\message{reflabel \string#1 is undefined.}%
\nref#1{need to supply reference \string#1.}\fi%
\vphantom{\hphantom{#1}}\edef\next{#1}\ifx\next\em@rk\def\next{}%
\else\ifx\next#1\ifodd\count255\relax\xref#1\count255=0\fi%
\else#1\count255=1\fi\let\next=\r@fs\fi\next}
%

%
\newwrite\ffile\global\newcount\figno \global\figno=1
\def\fig{Fig.~\the\figno\nfig}
\def\nfig#1{\xdef#1{Fig.~\the\figno}%
\writedef{#1\leftbracket Fig.\noexpand~\the\figno}%
\ifnum\figno=1\immediate\openout\ffile=figs.tmp\fi\chardef\wfile=\ffile%
\immediate\write\ffile{\noexpand\medskip\noexpand\item{Fig.\ \the\figno. }
\reflabeL{#1\hskip.55in}\pctsign}\global\advance\figno by1\findarg}
\def\xfig{\expandafter\xf@g}\def\xf@g Fig.\penalty\@M\ {}
\def\figs#1{Figs.~\f@gs #1{\hbox{}}}
\def\f@gs#1{\edef\next{#1}\ifx\next\em@rk\def\next{}\else
\ifx\next#1\xfig #1\else#1\fi\let\next=\f@gs\fi\next}
\newwrite\lfile
{\escapechar-1\xdef\pctsign{\string\%}\xdef\leftbracket{\string\{}
\xdef\rightbracket{\string\}}\xdef\numbersign{\string\#}}

\def\writestop{\def\writestoppt{\immediate\write\lfile{\string\pageno%
\the\pageno\string\startrefs\leftbracket\the\refno\rightbracket%
\string\def\string\secsym\leftbracket\secsym\rightbracket%
\string\secno\the\secno\string\meqno\the\meqno}\immediate\closeout\lfile}}
\def\writestoppt{}\def\writedef#1{}
\def\seclab#1{\xdef #1{\the\secno}\writedef{#1\leftbracket#1}\wrlabeL{#1=#1}}
\def\subseclab#1{\xdef #1{\secsym\the\subsecno}%
\writedef{#1\leftbracket#1}\wrlabeL{#1=#1}}
\newwrite\tfile \def\writetoca#1{}
\def\leaderfill{\leaders\hbox to 1em{\hss.\hss}\hfill}
\def\writetoc{\immediate\openout\tfile=toc.tmp
   \def\writetoca##1{{\edef\next{\write\tfile{\noindent ##1
   \string\leaderfill {\noexpand\number\pageno} \par}}\next}}}

%
\catcode`\@=12 
%
\edef\tfontsize{\ifx\answ\bigans scaled\magstep3\else scaled\magstep4\fi}
\font\titlerm=cmr10 \tfontsize \font\titlerms=cmr7 \tfontsize
\font\titlermss=cmr5 \tfontsize \font\titlei=cmmi10 \tfontsize
\font\titleis=cmmi7 \tfontsize \font\titleiss=cmmi5 \tfontsize
\font\titlesy=cmsy10 \tfontsize \font\titlesys=cmsy7 \tfontsize
\font\titlesyss=cmsy5 \tfontsize \font\titleit=cmti10 \tfontsize
\skewchar\titlei='177 \skewchar\titleis='177 \skewchar\titleiss='177
\skewchar\titlesy='60 \skewchar\titlesys='60 \skewchar\titlesyss='60
\def\titlefont{\def\rm{\fam0\titlerm}
\textfont0=\titlerm \scriptfont0=\titlerms \scriptscriptfont0=\titlermss
\textfont1=\titlei \scriptfont1=\titleis \scriptscriptfont1=\titleiss
\textfont2=\titlesy \scriptfont2=\titlesys \scriptscriptfont2=\titlesyss
\textfont\itfam=\titleit \def\it{\fam\itfam\titleit}\rm}
 \ifx\answ\bigans\else scaled\magstep1\fi
\ifx\answ\bigans\def\abstractfont{\tenpoint}\else
\font\abssl=cmsl10 scaled \magstep1
\font\absrm=cmr10 scaled\magstep1 \font\absrms=cmr7 scaled\magstep1
\font\absrmss=cmr5 scaled\magstep1 \font\absi=cmmi10 scaled\magstep1
\font\absis=cmmi7 scaled\magstep1 \font\absiss=cmmi5 scaled\magstep1
\font\abssy=cmsy10 scaled\magstep1 \font\abssys=cmsy7 scaled\magstep1
\font\abssyss=cmsy5 scaled\magstep1 \font\absbf=cmbx10 scaled\magstep1
\skewchar\absi='177 \skewchar\absis='177 \skewchar\absiss='177
\skewchar\abssy='60 \skewchar\abssys='60 \skewchar\abssyss='60
\def\abstractfont{\def\rm{\fam0\absrm}
\textfont0=\absrm \scriptfont0=\absrms \scriptscriptfont0=\absrmss
\textfont1=\absi \scriptfont1=\absis \scriptscriptfont1=\absiss
\textfont2=\abssy \scriptfont2=\abssys \scriptscriptfont2=\abssyss
\textfont\itfam=\bigit \def\it{\fam\itfam\bigit}\def\footnotefont{\tenpoint}%
\textfont\slfam=\abssl \def\sl{\fam\slfam\abssl}%
\textfont\bffam=\absbf \def\bf{\fam\bffam\absbf}\rm}\fi
\def\tenpoint{\def\rm{\fam0\tenrm}
\textfont0=\tenrm \scriptfont0=\sevenrm \scriptscriptfont0=\fiverm
\textfont1=\teni  \scriptfont1=\seveni  \scriptscriptfont1=\fivei
\textfont2=\tensy \scriptfont2=\sevensy \scriptscriptfont2=\fivesy
\textfont\itfam=\tenit \def\it{\fam\itfam\tenit}\def\footnotefont{\ninepoint}%
\textfont\bffam=\tenbf \def\bf{\fam\bffam\tenbf}\def\sl{\fam\slfam\tensl}\rm}
\font\ninerm=cmr9 \font\sixrm=cmr6 \font\ninei=cmmi9 \font\sixi=cmmi6
\font\ninesy=cmsy9 \font\sixsy=cmsy6 \font\ninebf=cmbx9
\font\nineit=cmti9 \font\ninesl=cmsl9 \skewchar\ninei='177
\skewchar\sixi='177 \skewchar\ninesy='60 \skewchar\sixsy='60
\def\ninepoint{\def\rm{\fam0\ninerm}
\textfont0=\ninerm \scriptfont0=\sixrm \scriptscriptfont0=\fiverm
\textfont1=\ninei \scriptfont1=\sixi \scriptscriptfont1=\fivei
\textfont2=\ninesy \scriptfont2=\sixsy \scriptscriptfont2=\fivesy
\textfont\itfam=\ninei \def\it{\fam\itfam\nineit}\def\sl{\fam\slfam\ninesl}%
\textfont\bffam=\ninebf \def\bf{\fam\bffam\ninebf}\rm}
%
%

\hyphenation{anom-aly anom-alies coun-ter-term coun-ter-terms}
\def\inv{^{\raise.15ex\hbox{${\scriptscriptstyle -}$}\kern-.05em 1}}

\def\Dsl{\,\raise.15ex\hbox{/}\mkern-13.5mu D} 
\def\dsl{\raise.15ex\hbox{/}\kern-.57em\partial}

\def\tr{{\rm tr}} 
\font\bigit=cmti10 scaled \magstep1
\def\lspace{\ifx\answ\bigans{}\else\qquad\fi}
\def\lbspace{\ifx\answ\bigans{}\else\hskip-.2in\fi} 
\def\boxeqn#1{\vcenter{\vbox{\hrule\hbox{\vrule\kern3pt\vbox{\kern3pt
	\hbox{${\displaystyle #1}$}\kern3pt}\kern3pt\vrule}\hrule}}}
\def\mbox#1#2{\vcenter{\hrule \hbox{\vrule height#2in
		\kern#1in \vrule} \hrule}}  
%

\def\psibar{\overline\psi}

\def\darr#1{\raise1.5ex\hbox{$\leftrightarrow$}\mkern-16.5mu #1}

\def\roughly#1{\raise.3ex\hbox{$#1$\kern-.75em\lower1ex\hbox{$\sim$}}}

\def\prp#1#2#3{Phys. Rep. {\bf #1} (19#2) #3}
\def\prl#1#2#3{Phys. Rev. Lett. {\bf #1} (19#2) #3}
\def\jpl#1#2#3{JETP Lett. {\bf #1} (19#2) #3}
\def\snp#1#2#3{Sov. J. Nucl. Phys. {\bf #1} (19#2) #3}
\def\pra#1#2#3{Phys. Rev. {\bf A#1} (19#2) #3}
\def\prb#1#2#3{Phys. Rev. {\bf B#1} (19#2) #3}
\def\prd#1#2#3{Phys. Rev. {\bf D#1} (19#2) #3}
\def\ijm#1#2#3{Int. J. Mod. Phys. {\bf A#1} (19#2) #3}
\def\zps#1#2#3{Z. Phys. {\bf #1} (19#2) #3}
\def\prv#1#2#3{Phys. Rev. {\bf #1} (19#2) #3}
\def\npb#1#2#3{Nucl. Phys. {\bf B#1} (19#2) #3}
\def\cmp#1#2#3{Commun. Math. Phys. {\bf #1} (19#2) #3}
\def\jpc#1#2#3{J. Phys. {\bf C#1} (19#2) #3}
\def\plb#1#2#3{Phys. Lett. {\bf B#1} (19#2) #3}
\def\aps#1#2#3{Ann. Phys. {\bf #1} (19#2) #3}
\def\mpl#1#2#3{Mod. Phys. Lett. {\bf A#1} (19#2) #3}
\def\zpc#1#2#3{Z. Phys. {\bf C#1} (19#2) #3}
\def\ncm#1#2#3{Nuovo Cim. {\bf #1} (19#2) #3}
\def\rmp#1#2#3{Rev. Mod. Phys. {\bf #1} (19#2) #3}
\def\tmp#1#2#3{Theor. Math. Phys. {\bf #1} (19#2) #3}
\def\pla#1#2#3{Phys. Lett. {\bf A#1} (19#2) #3}

\def\ay{\varepsilon}
\def\ks{\kappa}
\def\jv{\upsilon}
\def\kv{\iota}
\def\by{\varsigma}

\def\cc{c}

\Title{\vbox{\baselineskip12pt\hbox{UB-ECM-PF 95/16}
\hbox{CCNY-HEP 96/11}
}}
{\vbox{\centerline{Exact renormalization group study of
fermionic theories$^*$}}}\footnote{}{$^*$\ This work is supported
in part by funds provided
by the M.E.C. under contracts AEN95-0590 and SAB94-0087.}
\centerline{Jordi Comellas\footnote{$^a$}
{E-mail: {\tt comellas@sophia.ecm.ub.es}}}
\medskip
\centerline{\it Departament d'Estructura i Constituents de la Mat\`eria}
\centerline{\it Facultat de F\'\i sica, Universitat de Barcelona}
\centerline{\it Diagonal~647,~08028 Barcelona, Spain}
\bigskip
\centerline{Yuri Kubyshin\footnote{$^b$}
{E-mail: {\tt kubyshin@theory.npi.msu.su}}}
\medskip
\centerline{\it Institute of Nuclear Physics}
\centerline{\it Moscow State University}
\centerline{\it 119899 Moscow, Russia}
\bigskip
\centerline{Enrique Moreno\footnote{$^c$}
{E-mail: {\tt moreno@scisun.sci.ccny.cuny.edu}}}
\medskip
\centerline{\it Department of Physics}
\centerline{\it City College of New York}
\centerline{\it New York, NY 10031, U.S.A.}

\vskip .3in

The exact renormalization group approach (ERG) is developed 
for the case of pure
fermionic theories by deriving a Grassmann version of the ERG
equation and applying it to the study
of fixed point
solutions and critical exponents of the two-dimensional
chiral Gross-Neveu model.  An approximation based on the derivative
expansion and a further truncation in the number of fields is used.
Two solutions
are obtained analytically in the limit $N\to \infty $, with $N$ being the
number of fermionic species.
For finite $N$ some fixed point
solutions, with their anomalous dimensions and critical exponents,
are computed numerically.  The issue of separation of physical results from
the numerous spurious ones is discussed.  We argue that one of the
solutions we find can be identified with that of Dashen and Frishman,
whereas the others seem to be new ones.

\Date{08/96}

\newsec{Introduction}

One of the major issues in QFT is the search
for non-perturbative results.  In particular little is known about the phase
structure (fixed points, critical exponents, etc.) of most
physically interesting
theories.  Nor have intriguing questions about the renormalization flows,
like the proper extension of the c-theorem
\ref\Zam{A.B. Zamolodchikov, \jpl{43}{86}{730}; \snp{46}{87}{1090}},
been completely understood
\ref\cth{J.L. Cardy, \plb{215}{88}{749}\semi H. Osborn,
\plb{222}{89}{97}\semi I. Jack and H. Osborn, \npb{343}{90}{647}\semi
A. Cappelli, D. Friedan and J.I. Latorre, \npb{352}{91}{616}\semi A. Cappelli,
J.I. Latorre and X. Vilas\'\i s-Cardona, \npb{376}{92}{510}}.

\nref\Wil{K.G. Wilson, \prb{4}{71}{3174}; 3184}
\nref\WK{K.G. Wilson and J. Kogut, \prp{12}{74}{75}}
\nref\Fisher{K.G. Wilson, \prl{28}{72}{548}\semi
M.E. Fisher and K.G. Wilson, \prl{28}{72}{240}\semi M.E. Fisher
\rmp{46}{74}{597}}
\nref\Pol{J. Polchinski, \npb{231}{84}{269}}
\nref\pert{J. Feldman, J. Magnen,
V. Rivasseau and R. S\'ene\'or, \cmp{99}{85}{273}\semi
K. Gawedzki and A. Kupiainen, \cmp{99}{85}{197}}
\nref\pertgauge{
B.J. Warr, \aps{183}{88}{1};59\semi
G. Keller and C. Kopper, \plb{273}{91}{323}\semi
C. Becchi, ``On the construction of renormalized quantum field theory
using renormalization group techniques'', in ``Elementary particles, Field
theory and Statistical mechanics'', ed. by M. Bonini, G. Marchesini
and E. Onofri,
(Parma University, Parma, 1993)}
\nref\HH{A. Hasenfratz and P. Hasenfratz, \npb{270}{86}{687}}
One of the methods capable of handling such problems
is the exact renormalization group (ERG, hereafter).
Originally developed by Wilson in his seminal articles in the early seventies
\Wil\ (see Ref.~\WK\ for a classical
review), it has recently attracted much attention.
Although firstly used
for studies of critical phenomena in condensed matter problems \Fisher\
its scope includes many other fields like particle
theory, as it was demonstrated by Polchinski in an elegant paper where
he proved the perturbative renormalizability of $\lambda \phi^4$ theory in a
quite simple way
\Pol.
Similar manipulations have led to quite interesting results regarding the
study of different
aspects of the perturbation expansion around a Gaussian fixed point and its
associated diagramatic expansion
\refs{\pert, \pertgauge}.
However, the power of the ERG relies on the possibility of obtaining
quantitative knowledge about
the renormalization flows, in particular one piece of information which is
probably the most valuable, due to its universality: the critical
exponents
\HH.

\nref\weq{F.J. Wegner and A. Houghton, \pra{8}{73}{401}}
\nref\wegner{F.J. Wegner, \jpc{7}{74}{2098}}
\nref\weinberg{S. Weinberg in ``Understanding the Fundamental Constituents
of Matter'', Erice 1976, ed. A. Zichichi (Plenum Press, New York, 1978)}
\nref\nond{J.F. Nicoll, T.S. Chang and H.E. Stanley, \pla{57}{76}{7}\semi
V.I. Tokar, \pla{104}{84}{135}\semi
C. Wetterich, \npb{352}{91}{529}\semi
A.E. Filipov and S.A. Breus, \pla{158}{91}{300}\semi
M. Alford, \plb{336}{94}{237}\semi
U. Ellwanger, \zpc{62}{94}{503}\semi
S. Bornholdt, N. Tetradis and C. Wetterich, \plb{348}{95}{89}\semi
D. Litim and N. Tetradis,
``Analytical Solutions of the Exact Renormalization Group Equations'',
hep-th/9501042}
\nref\intj{T.R. Morris, \ijm{9}{94}{2411}}
\nref\der{
J.F. Nicoll, T.S. Chang and H.E. Stanley, \prl{33}{74}{540}\semi
G.R. Golner, \prb{33}{86}{7863}\semi
G. Felder, \cmp{111}{87}{101}\semi
P. Hasenfratz and J. Nager, \zpc{37}{88}{477}\semi
C. Bagnuls and C. Bervillier, \prb{41}{90}{402}\semi
A.E. Filippov and A.V. Radievsky, \pla{169}{92}{195}\semi
S.-B. Liao and J. Polonyi, \aps{222}{93}{122}; \prd{51}{95}{4474}\semi
N. Tetradis and C. Wetterich, \npb{422}{94}{541}\semi
T.R. Morris, \plb{345}{95}{139};
``Momentum Scale
Expansion of Sharp Cutoff Flow Equations'', hep-th/9508017\semi
K. Halpern and K. Huang, \prl{74}{95}{3526}}
\nref\redtwo{T.R. Morris, \plb{329}{94}{241}}
\nref\spura{P.E. Haagensen, Y. Kubyshin, J.I. Latorre and E. Moreno,
\plb{323}{94}{330}}\nref\spurb{T.R. Morris, \plb{334}{94}{355}}
\nref\BHLM{R.D. Ball, P.E. Haagensen, J.I. Latorre and E. Moreno,
\plb{347}{95}{80}}
\nref\finT{N. Tetradis and C. Wetterich, \npb{398}{93}{659};
\ijm{9}{94}{4029}\semi
M. Reuter, N. Tetradis and C. Wetterich, \npb{401}{93}{567}\semi
S.-B. Liao and M. Strickland, ``Renormalization group approach to
field theory at finite temperature'', hep-th/9501137\semi
S. Bornholdt and N. Tetradis, ``High temperature phase transition in two scalar
theories'', hep-ph/9503282}
\nref\gauge{
M. Reuter and C. Wetterich, \npb{417}{94}{181}; {\bf B427} (1994) 291\semi
U. Ellwanger, \plb{335}{94}{364}\semi
M. Bonini, M. D'Attanasio and G. Marchesini, \npb{437}{95}{163};
\plb{346}{95}{87}\semi
T.R. Morris, \plb{357}{95}{225}; ``Two Phases for Compact $U(1)$ Pure Gauge
Theory in Three Dimensions'', hep-th/9505003\semi
U. Ellwanger, M. Hirsch and A. Weber, ``Flow equations for the relevant
part of the pure Yang-Mills action'', hep-th/9506019\semi
S.-B. Liao, ``Operator cutoff regularization and renormalization group
in Yang-Mills theory'', hep-th/9511046}
\nref\yuk{
A. Margaritis, G. \'Odor and A. Patk\'os, \zpc{39}{88}{109}\semi
M. Maggiore, \zpc{41}{89}{687}\semi
U. Ellwanger and L. Vergara, \npb{398}{93}{52}\semi
T.E. Clark, B. Haeri and S.T. Love, \npb{402}{93}{628}\semi
U. Ellwanger and C. Wetterich, \npb{423}{94}{137}\semi
T.E. Clark, B. Haeri, S.T. Love, M.A. Walker and W.T.A. ter Veldhuis,
\prd{50}{94}{606}\semi
D.U. Jungnickel and C. Wetterich, ``Effective Action for the Chiral Quark Meson
Model'', hep-ph/9505267}
\nref\QED{M. Bonini, M. D'Attanasio and G. Marchesini, \npb{418}{94}{81};
\plb{329}{94}{249}}
The ERG approach is based on writing down a functional differential equation
that expresses
how the action changes when we integrate out high energy
modes.  This is the so-called ERG equation, and it is the cornerstone of the
whole technique: with it and together with
the most general action consistent with
the symmetries of the model,
the complete set of
$\beta $-functions
can be computed, and from these the location of the fixed points and their
exponents.
However, because of practical reasons it is impossible to handle all
possible
operators which could be included into the action.
One must choose a more selective
criterion, rather than simply to be consistent with the demanded symmetries,
that is, one must choose a reasonable truncation of the general
expansion.  Usual approximations attempt
to restrict the space of interactions to a
reasonable number of operators,
e.g.~replacing the effective action for a non-derivative
effective potential or expanding the action in powers of the momentum.
For instance, to study the Wilson fixed point of $\lambda \phi^4$
in three dimensions, the authors of Ref.~\HH\ consider only arbitrary
polynomials in the fields without derivatives, but the same type of problem
has also been addressed by
changing the ERG equation and/or by
considering other types of truncations \refs{\WK, \weq-\BHLM}.

\nref\fermGN{K. Gawedzki and A. Kupiainen, \prl{54}{85}{2191};
{\bf 55} (1985) 363;
\cmp{102}{85}{1}; \npb{262}{85}{33}}
\nref\Fermi{E. Fermi,
\zps{88}{34}{161}}
\nref\QCD{J. Bijnens, C. Bruno and E. de Rafael, \npb{390}{93}{501}\semi
A.A. Andrianov and V.A. Andrianov \tmp{94}{93}{3}}
\nref\NJL{Y. Nambu and G. Jona-Lasinio, \prv{122}{61}{345}}
\nref\top{A. Hasenfratz, P. Hasenfratz, K. Jansen, J. Kuti and
Y. Shen, \npb{365}{91}{79}\semi J. Zinn-Justin, \npb{367}{91}{105}}
\nref\tech{V.A. Miransky, M. Tanabashi and K. Yamawaki,
\plb{221}{89}{177};
\mpl{4}{89}{1043}\semi W.A. Bardeen, C.T. Hill and M. Lindner,
\prd{41}{90}{1647}}
\nref\stmod{A. Hasenfratz and P. Hasenfratz, \plb{297}{92}{166}}
ERG methods have been used in more complicated cases, like
phase transitions at finite temperature \finT,
theories with gauge interactions \refs{\pertgauge, \gauge}
and theories with
fermions as fundamental particles.
In the last case, it usually relies on a certain kind of bosonization,
usually consisting in coupling fermionic bilinears with scalar fields, and
then studying self-interactions of these scalar fields\foot{The main
exceptions are Ref.~\QED\ which deals with perturbative properties
of the Gaussian fixed point and the series of Ref.\fermGN\ which lead
to the proof of the existence of the Gross-Neveu model in $2+\epsilon$
dimensions.  We are seeking, however,
non-perturbative quantitative information out of
the ERG, rather than rigorous formal statements.} \yuk.
We feel,
nonetheless, that this approach is
unsatisfactory.  The reason is that one should
learn how to deal with pure spinor theories without simplifying the problem
to a scalar one.
Moreover, there are some quite
interesting
phenomenological models described entirely with spin $1/2$ fields, like the
celebrated Fermi theory of weak interactions \Fermi,
models for resonance physics \QCD\
based on extensions of the
Nambu-Jona-Lasinio action \NJL,
or even models explaining the symmetry-breaking
sector of the electroweak theory, especially in connection with
technicolor theories\foot{Note, however,
Ref.~\top}
\tech.
It is also compelling the curiosity that all fields in the Standard Model
can be expressible as fermionic fields \stmod.
Surely, further extensions have to be dealt with afterwards, like
scalar and spin $1/2$ fields coupled together and spinor particles interacting
through gauge fields.

On the other hand, we remind the reader that fermions are
not always easily manageable by non-perturbative methods,
e.g.~Lattice Field Theory.  On the contrary, we will show that,
once truncated, the
ERG equation treats fermions and bosons very similarly, thus making possible
a nearly immediate translation of knowledge from one case to the other.

Our purpose is to
work with a sample theory based solely on spinor fields and to develop a method
of obtaining numerical non-perturbative information from it (e.g.~some
critical exponents).
With this motivation an ERG equation is derived, similar to the
bosonic one by Polchinski \Pol, and applied to
a particular model.  We try to emphasize
throughout the article that many of the peculiarities encountered are
nothing more than the translation of their counterparts
already found in previous papers on the ERG for bosonic theories.

To begin with, one may choose an appropriate model, relatively simple and
non-trivial.  As usual, the two-dimensional world is a perfect site
where to look
for. Indeed, the two-dimensional Clifford algebra is the simplest
one, generated by the
well-known Pauli matrices.

Moreover,
the study of self-interacting fermionic theories in $d=2$ can be traced
back to the work of Thirring \ref\Thir{W. Thirring, \aps{3}{58}{91}},
where he proposed a massless model of a single Fermi field containing a
quartic self-interaction.  It can be solved exactly \ref\texac{V.
Glaser, \ncm{9}{58}{990}\semi K. Johnson, \ncm{20}{61}{773}\semi C.
Sommerfield, \aps{26}{63}{1}\semi B. Klaiber, in "Lectures in
Theoretical Physics'', ed. by A.O. Barut and W.E. Brittin, (Gordon and
Breach, New York, 1968)} and it presents some interesting features,
most of
them probably unexpected.  It is perturbatively renormalizable, as power
counting arguments suggest, but what is not expected from naive
arguments is that it does not describe an interacting theory, but a
trivial one \ref\SG{S. Coleman, \prd{11}{75}{2088}}.

The model becomes non-trivial when $N$ species
of fermions transforming under a global representation of the
unitary group $U(N)$ are
considered. This
is the Gross-Neveu model \ref\GN{D. Gross and A. Neveu,
\prd{10}{74}{3235}}, which is asymptotically free and renormalizable,
within perturbation theory and also within the $1/N$ expansion.  However,
none of these approximations is capable to find any
non-trivial fixed point for $d=2$.
(Actually the $1/N$ expansion shows
a non-trivial fixed point of order $d-2$
in $d$ dimensions).

An interesting modification leads to the so-called chiral Gross-Neveu
model \GN,
which is chosen
to have the additional symmetry of the $U_R(N)\times U_L(N)$ chiral group.
As in the previous case no fixed points, besides the Gaussian one,
can be found within
the $1/N$ approximation.  Nevertheless, other non-perturbative
techniques are available.  Thus the
quartic interaction of the chiral Gross-Neveu model
can be expressed, after a Fierz
transformation, as a current-current interaction and the latter
allows an operator analysis of the model within the current algebra
approach.
Such a study is carried out in Ref.~\ref\DF{R.
Dashen and Y. Frishman, \plb{46}{73}{439};
\prd{11}{75}{2781}} where, exploiting conformal
techniques, two critical curves in the space of couplings are found for which
the theory is scale invariant.  One of the lines corresponds to
the abelian Thirring model,  whereas the other one is truly non-trivial
and does not
pass through the origin.  A very remarkable fact is that this result is
exact and  is not given  by a zero of a $\beta$-function,
neither in perturbation theory nor in the large $N$ expansion.  For this
continuum set of critical theories the value of the coupling constant,
associated to the abelian degrees of freedom, is arbitrary while the
coupling associated to the $SU(N)$ degrees of freedom is fixed to
be equal to zero
or $4 \pi /(N+1)$.  It is important to notice, however, that the fixed
point is not unique and, as it happens in the Thirring model too,
depends on
an arbitrary parameter related to regularization ambiguities.

More recently, using bosonization, current algebra and conformal
techniques, other non-trivial fixed points in two-dimensional
fermionic models were found \ref\ofp{P.K. Mitter and P.H. Weisz,
\prd{8}{73}{4410}\semi
E. Moreno and F. Schaposnik, \ijm{4}{89}{2827}\semi C. Hull and O.A.
Soloviev, ``Conformal Points and Duality of Non-Abelian Thirring Models
and Interacting WZNW Models", hep-th/9503021}.  However a lot of work
has to be done in order to gain a clearer understanding of
the complete phase diagram.

Before ending this section let us describe shortly our results.
We studied the chiral Gross-Neveu model with a Polchinski-type
ERG equation and we projected the space
of local interactions onto the 106-dimensional subspace generated by
terms with at most six fermions and three derivatives.

Within the large $N$ approximation we found two different non-trivial
$N\to \infty$ limits, leading to qualitatively different results.
One of them leads to a continuous family of fixed points along the
direction of the $U(1)$ excitations, similar to that of Dashen and
Frishman. However, unlike their case, the anomalous dimension $\eta$ vanishes
at leading order in $1/N$. The critical exponents can be computed
analytically and most of them coincides with the canonical values.
However for a wide class of regularization schemes the most relevant
critical exponent is non-trivial and takes the value
$\lambda_1=1.1231...$.
The other solution gives a non-trivial (but scheme dependent) anomalous
dimension ($\eta = 1.0 \sim 1.4$ in the range studied), but the fixed
point is isolated. The relevant critical exponent in the same range
is in this case $\lambda_1 = 2.18 \sim 2.26$.

For finite $N$ the fixed point equations can be solved only numerically.
We found a plethora of solutions, most of them spurious due to the
nature of the approximation. As we will discuss later, it is very
difficult to discriminate between good and fictitious solutions. We
took specially care of those solutions that for large $N$
match the $1/N$ fixed points. For both cases the solutions are strongly
dependent on the scheme though the dependence can be mitigated using a
minimum sensitivity criterion. In one case
the solution is defined for any value of $N$, with
$N\ \eta$ growing asymptotically to $4.9$, and
$\lambda_1$ decreases to its $1/N$ limit $1.123...$.
In the other case the solution disappears unexpectedly at $N\approx 142$
where it merges, as we will carefully describe,
with another kind of solution. At the bifurcation point we have
$\eta=1.88$ and $\lambda_1=5.8$.

Finally we analyzed the case $N=1$. This case has to be treated
separately because in absence of flavour the Fierz transformations
impose additional constraints that reduce considerably the total number
of independent operators. The results in this case are disappointing:
the fixed point solution is isolated and not a continuous family as in
the Thirring model. This property is fulfilled in the previous order
approximation (terms with less than three derivatives) and it is unclear
for us why it is lost at this order. For this case the values of the
anomalous dimension has an extremely wide variation with the
renormalization scheme ($\eta =1 \sim 10$).

Let us remind the reader that in general the fixed-point actions are scheme
dependent and they contain
any possible operator, thus making it
very difficult to interpret them directly.
Nevertheless, any fixed point is characterized by that piece of information
that is universal and, thus, physically relevant.
This contains the number of relevant directions and their associate exponents.
For these reasons, we have refrained from presenting
in detail
the fixed-point actions but, rather, we have concentrated in discussing
the universal properties of our results.

The article is organized as follows.  In section 2 we derive an ERG
equation for pure fermionic theories in any dimension.  Sect.~3 is
devoted to the construction of the action.  The chiral Gross-Neveu model
is defined through its symmetries and the truncation we use is
explained.  The calculation of the $\beta$-functions, fixed points and
the corresponding critical exponents is divided into sections 4 and 5,
while the sixth one contains a summary of results and some conclusions.
Because of the unavoidable increasing complexity of the notation (we
will be dealing with 106 operators) we have included an Appendix (A)
where all our conventions are summarized.  Although any new symbol is
clearly defined when it appears, we thought it would be helpful to have
all them condensed in a single page.  We have also included one appendix
(B) to write down the whole action and another one (C) to present the
complete set of $\beta$-functions.

\newsec{ERG equation for fermionic theories}

In this section
we derive an ERG equation for a field
theory written in terms of spinor quantities, on a Euclidean space of
dimension $d$.
Its role is to
dictate the behaviour of
the action as we integrate out modes, that is, how the action of our
effective theory has to be modified when we vary the characteristic
scale (cut-off) $\Lambda $, while keeping the $S$-matrix elements invariant.
More concretely,
if we parametrize the renormalization flow with $t\equiv -\ln{\Lambda\over
\Lambda_0}$,
then the ERG equation
will provide us with
a sufficient condition to ensure $\dot Z=0$, 
$\Lambda_0$ being a fixed scale and
$Z$ the generating
functional for the connected Green's functions.

We will consider, thus,
a general theory whose action is expressible as a function
of spinor fields only\foot{We assume that
Weinberg's conjecture
\ref\Weinberg{S. Weinberg,
Physica {\bf 96A} (1979) 327} is valid: an arbitrarily general action
leads to arbitrarily general $S$-matrix elements and vice versa.}
and artificially split it into
\eqn\kinpint{S=S_{kin}+S_{int}\ .}  $S_{int}$ is
an arbitrary polynomial in the fields and momenta (we will work
always in momentum space) and $S_{kin}$ is a regulated version of the
usual kinetic term,
\eqn\fkt{S_{kin}=\int_p\psibar_{-p}P^{-1}_{\Lambda }\left( p\right)
\psi_p\ ,} where $\int_p$ stands for $\int d^dp$ and
$P_{\Lambda }$ is the matrix
\eqn\matp{P_{\Lambda }\left( p\right) =({2 \pi})^d {K_{\Lambda
}\left( p^2\right) \over p^2}i \rlap/p\ ,} with
$K_{\Lambda }\left( z\right) $ an analytic function over the whole finite
complex plane that vanishes faster than
any power-law when $z\to +\infty $ and is
normalized to be $K_{\Lambda }\left( 0\right) = 1$ \ref\Ball{R.D. Ball and
R.S. Thorne, \aps{236}{94}{117}}.

In the following, we will consider expectation values defined
by
\eqn\fvev{\left< X \right>
\equiv \int {\cal D} \psi {\cal D} \psibar \,\, X \, e^{ -S\left[ \psibar
,\psi ;\Lambda \right] +\int_p
\overline\chi_{-p} Q_{\Lambda }^{-1}\left( p\right)\psi_p+
\int_p \psibar_{-p}Q_{\Lambda }^{-1}\left( p\right)\chi_p+f_{\Lambda }}\ ,}
where $X$ is any operator, $\chi_p$, $\overline\chi_p$ are Grassmann sources,
$Q_{\Lambda }\left( p\right)$ is another regulating function with
analogous properties to $P_{\Lambda }\left( p\right) $
and, finally, $f_{\Lambda }$ is a c-number independent of the fields.

With the above conventions,
the starting point of the derivation is the observation that a
functional integral of a total functional derivative vanishes, which leads
us to
\eqn\fid{\eqalign{\biggl< \biggl( {\delta \over \delta \psi_p}
-\psibar_{-p}P^{-1}_{\Lambda } + \overline\chi_{-p}
Q^{-1}_{\Lambda }\biggr) \biggr. & \dot
P_{\Lambda }\biggl. \biggl( {\delta \over \delta
\psibar_{-p}}+P^{-1}_{\Lambda }\psi_p-Q^{-1}_{\Lambda
}\chi_p\biggr) \biggr>  \cr = - \Bigl< \tr \left(
P^{-1}_{\Lambda } \dot P_{\Lambda }
\right) \delta \left( 0\right) \Bigr> & - \left<
\left( \psibar_{-p}P^{-1}_{\Lambda }-\overline\chi_{-p}
Q^{-1}_{\Lambda }\right) \dot
P_{\Lambda
} \left( P^{-1}_{\Lambda }
\psi_p-Q^{-1}_{\Lambda }\chi_p\right) \right> ,\cr }}
where the trace is over spinor indices.  This is the counterpart of
Eq.~(1.9) of Ref.~\Ball\
and, as there, it can be used to identify the rate of change of
the kinetic term,
\eqn\evke{\eqalign{\int_p\left<
\psibar_{-p}\dot P^{-1}_{\Lambda }\psi_p\right> = &
\int_p\left<
{\delta S_{int}\over \delta \psi_p}\dot P_{\Lambda }
{\delta S_{int}\over \delta \psibar_{-p}}
- {\delta \over \delta \psi_p} \dot P_{\Lambda }
{\delta \over \delta \psibar_{-p}} S_{int}\right> \cr +
\int_p \left<
\psibar_{-p}P^{-1}_{\Lambda }\dot P_{\Lambda
}Q^{-1}_{\Lambda }\chi_p \right. & +
\left. \overline\chi_{-p}Q^{-1}_{\Lambda }\dot
P_{\Lambda }P^{-1}_{\Lambda }
\psi_p\right>
+ \int_p \left< \overline\chi_{-p}Q^{-1}_{\Lambda }
\dot P_{\Lambda }Q^{-1}_{\Lambda }
\chi_p\right. + \left. \tr \left( P^{-1}_{\Lambda }
\dot P_{\Lambda }\right) \delta \left( 0\right)
\right> .\cr }}

On the other hand, by imposing that the generating
functional is independent of the scale $\Lambda $ we find the
relation
\eqn\fcond{\eqalign{\left< \dot S_{int}\right> = & - \int_p \left<
\eta \, \psibar_{-p} P^{-1}_{\Lambda }\psi_p +
\psibar_{-p} \dot P^{-1}_{\Lambda }\psi_p\right> - {\eta \over 2}
\int_p \left< \psibar_p {\delta S_{int} \over \delta \psibar_p}
+ \psi_p {\delta S_{int} \over \delta \psi_p} \right> \cr + &
\int_p\left<
{\eta \over 2}\, \overline\chi_{-p} Q^{-1}_{\Lambda }
\psi_p + {\eta \over 2}\, \psibar_{-p} Q^{-1}_{\Lambda }
\chi_p + \overline\chi_{-p}\dot Q^{-1}_{\Lambda
}\psi_p+\psibar_{-p} \dot Q^{-1}_{\Lambda }\chi_p \right> + \left<
\dot f_{\Lambda }\right> ,\cr }} where the anomalous
dimension is defined to be
\eqn\fad{\dot\psi_p={\eta \over 2}\, \psi_p,\ \ \
\dot{\overline\psi}_p ={\eta \over 2}\, \psibar_p.}

We can now combine Eqs.~\evke\ and \fcond\
to write, after some straightforward algebra, an equation for $\left<
\dot S\right>$.
It will be satisfied if
\eqn\fsdot{\eqalign{\dot S= & \int_p\left( {\delta S\over
\delta \psi_p}\dot P_{\Lambda }\left( p\right){\delta S\over \delta
\psibar_{-p}}-{\delta \over \delta \psi_p}\dot P_{\Lambda }\left(
p\right) {\delta S\over \delta \psibar_{-p}}\right) \cr + & \int_p\left[
{\delta S\over \delta \psi_p} \left( \dot P_{\Lambda }\left( p\right)
P^{-1}_{\Lambda }\left( p\right) \right) \psi_p -
\psibar_{-p}\left( P^{-1}_{\Lambda }\left( p\right) \dot P_{\Lambda
}\left( p\right) \right) {\delta S\over \delta
\psibar_{-p}}\right] .\cr }}
Note that our claim is that if \fsdot\ holds for functionals,
as it stands, then a similar equation will hold for expectation
values, which, in turn, ensures that all the Green's functions are
invariant under the flow.  We have just found, therefore, the
sufficient condition we were looking for.

We still have
to take into account the
effects produced by the rescalings needed after any Kadanoff
type of change in order to complete a RG transformation.  We have
partially included them when we consider not bare fields but
renormalized ones with some anomalous
dimension $\eta $.  What is left over is just the canonical
evolution of all quantities.  To compute them in closed form the
easiest way is to write \fsdot\
after having rescaled all variables with the appropriate powers of
$\Lambda $ to make them dimensionless,
\eqn\feq{
\eqalign{\dot S = 2 ({2\pi})^d \int_p K'\left(
p^2 \right) & \left( {\delta S\over
\delta \psi_p}i \rlap/p{\delta S\over \delta
\psibar_{-p}} -{\delta \over \delta \psi_p}i\rlap/p
{\delta S\over \delta \psibar_{-p}}\right) \cr + d S
+ \int_p \left( {1-d+\eta \over 2}-2p^2{K'\left( p^2
\right) \over K\left( p^2\right) }\right) & \left(
\psibar_p{\delta S\over \delta \psibar_p}+\psi_p {\delta S\over
\delta \psi_p}\right) - \int_p \left( \psibar p^{\mu }
{\partial '\over \partial  p^{\mu }}{\delta S\over \delta
\psibar_p}+\psi_p p^{\mu }{\partial '\over \partial p^{\mu }}
{\delta S\over \delta \psi_p}\right) ,\cr }}
where the prime in
$\partial '/\partial p^{\mu }$ means that the derivative does
not act on the momentum conservation delta functions and
thus only serves to count the powers of momenta of
a given functional.

Note that, once the above ERG equation is derived,
it would be easy to derive a similar one
for a model involving Yukawa couplings just by combining the
present manipulations with that of, for instance,
Refs.~\refs{\Pol ,\Ball}.
(The resemblance of \feq\ and a Polchinski type equation for scalar
theories is pretty evident.)

To fully specify the evolution of our theory
under the RG flow, we have also to write down equations for the terms
containing $\chi_p$ and $\overline\chi_p$ and the term with
neither the sources nor the fields.  Although we will not
need them in our analysis, just for the sake of completeness
we present the expressions obtained without further
comments\foot{For
a discussion with respect to the scalar case see again Ref.~\Ball.}:
\eqn\other{
Q_{\Lambda }\left( p\right) =
P_{\Lambda }\left( p\right) \tilde Q\left( p^2\right) \ ,
\
f_{\Lambda }=
- \int dt \int_p \tilde Q^{-2}\left( p^2\right) \overline\chi_{-p}\dot
P^{-1}_{\Lambda }\left( p^2\right) \chi_p
\ ,} with $\tilde Q\left(
p^2\right) $ a scalar
function that evolves according to the equation $\dot {\tilde
Q}\left( p^2\right) ={\eta \over 2} \tilde Q\left( p^2\right) $.

As a final comment about our equation is that we present it on Euclidean
space as it is customary in the field.  For our purposes, however, there is
nothing special about the Euclidean formulation, as finally what one obtains
is just a set of relations among coupling constants.  In fact,
we have also derived the counterpart of Eq.~\feq\ for
Minkowski space.  It is not so nice because of the extra presence of an
imaginary unit coming from the functional derivatives of the Minkowskian
``Boltzmann" factor $e^{i{\cal S}}$ in the second term.  Nevertheless,
with this equation we have computed the $\beta$-functions for a
simplified action (one without operators with six fields) in much the
same way we will explain later for Euclidean space: they are finite,
real and consistent with the desired symmetries, as they should be.
We have not proceeded further, but the parallelism between them and their
Euclidean counterparts strongly
supports the common lore
that both should contain the same physical
information and that the choice of space is much a matter of taste.
Nevertheless,
it would probably be nice to afford a complete calculation in Minkowski space.

\newsec{The action}

In this section we begin the discussion of an explicit example.
We first define it through its symmetries, then justify how one
can truncate its general action while still retaining non-trivial
information and, finally, we give the prescriptions we have actually used to
build it systematically.

The sample model is
that with $N$ spin $1/2$ two-dimensional Euclidean
fields that obey the discrete symmetries of parity, charge conjugation
and, to obtain reflection positive Green's functions, reflection
hermiticity
(see Ref.~\ref\book{J. Zinn-Justin, "Quantum
Field Theory and Critical Phenomena'', (Oxford Science Publications, Oxford,
1989)} for a precise definition of them).
We further impose
the continuous symmetries of Euclidean invariance and the chiral symmetry
$U(N)_R\times U(N)_L$.

For the definition to be consistent,
one has to check that the above classical symmetries of
the action will survive after quantization.  That is, one has to
ensure that the symmetries will be satisfied at any point of the flow if
they are satisfied by the initial conditions.  In our case
this is verified nearly immediately by just looking at Eq.~\feq . The
point is that the Kadanoff terms, which are the eventually dangerous
ones, essentially take the form, in spinor and flavour indices, of the
free kinetic term of the action.

The next step is to choose an appropriate truncation.  One would desire
a kind of derivative expansion, at least because it is quite efficient
when applied to bosonic theories \refs{\der, \BHLM}.
However, the similarity with the
scalar case cannot be carried that far.
The first important difference is that,
unlike the scalar case, the zero momentum approximation (effective potential)
is not feasible and the leading order is one with zero and one
derivative terms.
The reason is almost evident:
Eq.~\feq\ contains, due to the sum over polarizations, a $\rlap /p$ factor
in the Kadanoff terms, while a similar equation for bosons does not\foot{See,
for instance, Eq.~(18) of Ref.~\Pol.}.

Another significant difference is that in the scalar case a general
potential contains an infinite number of independent
functionals, whereas for finite $N$
a general product of fermionic fields with fixed number of
derivatives has in any case a finite number of terms due
to the statistics.  It is impossible to put twice the same Grassmann
quantity at the same point. Thus, for the fermionic case the derivative
expansion leads unavoidably to a polynomial approximation.
This has practical
consequences: the ERG equation becomes a large system of coupled non-linear
ordinary differential equations instead of a small set of coupled
partial differential equations, and the techniques to obtain
numerical results are different.
Furthermore, the number of different structures for an arbitrary large
value of $N$ grows extremely fast as the order of the derivatives
increases. In practice, it becomes practically intractable at order
$3$ unless the degree of the polynomial of the fields is also
truncated.
For this reason we work, up to a finite number of derivatives and
{\it also} up to a finite number of fields.
The remaining decision is to choose where to truncate.

We require that
a {\it sine qua non} property of a decent approximation is to allow
a nontrivial anomalous dimension.  Therefore,
we will keep as many
derivatives as needed to allow for a non-zero $\eta $,
within a reasonable number of fields (of the order of, say, twice the
number of derivatives).  With this criterion it is easy to realize that
one derivative and four fields do not work:  only the Gaussian fixed
point is obtained, with classical critical exponents.  Two derivatives
seem in principle sufficient.  However, once the $\beta$-functions are
obtained, it can be shown that the result $\eta=0$ is unavoidable,
thus forcing us to work with
terms up to three derivatives and six fields.

The final preparatory step is to write down the action.  To construct it
systematically we list all symmetries and study the restrictions
imposed by each of them.

We will work with the momentum representation and, in order to simplify
the notation, we will take the convention that any product of fields should
be eventually integrated over
the momentum carried by each field, with a common momentum conservation
delta function.  This would correspond to an integral over the whole
space of a product of fields and their derivatives (of any order) at the
same point.

i) $\underline{U(1)}$.  We begin with $U(1)$,
fermion number conservation.  Its consequences are well known: the
action must be built up of operators of the form
\eqn\genoper{S_{12}^{ab}\equiv
\psibar^a\left(p_1\right)\psi^b\left(p_2\right),\
\ \
P_{12}^{ab}\equiv \psibar^a\left(p_1\right)\gamma_s\psi^b\left(p_2\right),\
\ \
V_{12}^{j,ab}\equiv \psibar^a\left(p_1\right)\gamma^j\psi^b\left(p_2\right),
}
where we work in the momentum representation, $a$, $b$ denote flavour
indices and from now on the subindices of $S$, $P$, $V$ label the
fermion momenta.  The Clifford algebra is defined by
$\left\{\gamma^i,\gamma^j\right\}=2\delta^{ij}$ with
$\gamma_s=-i\gamma^1\gamma^2$. Note that in two dimensions there are no
other spinorial
structures, since $\gamma_s\gamma^j=i\epsilon^{jk}\gamma^k$.

ii) $\underline{Euclidean\ invariance}$.
The Euclidean invariance is also easily taken into account: one has only
to make sure that all Euclidean indices
are properly contracted.

iii) $\underline{SU(N)}$.
The next one is the
(vector) $SU(N)$ group.
If the fields transform under the fundamental representation,
all possible scalar operators can be
classified with the aid of Fierz reorderings.  In fact, it is
not difficult to show by means of Fierz transformations that a general local
operator
in the trivial representation, built from products of fermionic fields,
can be factored in terms of
\eqn\avop{S_{12}\equiv  S_{12}^{aa},\ \ \ P_{12}\equiv
P_{12}^{aa},\
\ \ V^j_{12}\equiv V^{j,aa}_{12}.}
Thus,
the simplest manner to get rid of the
internal group indices is to work with a basis written as products
of scalar, pseudoscalar and vector operators ($S$, $P$, $V^j$),
transforming under the trivial representation,
and powers of momenta.
Therefore, the simplicity of two dimensions has come to help
us again: a general functional
can be written in terms of only three ``building blocks", and
momenta.

iv) $\underline{SU_R(N)\times SU_L(N)}$.
To enlarge $SU(N)$ to $SU_R(N)\times SU_L(N)$ we realize
that the chiral invariant operators are constructed from the
combinations
\eqn\chir{V_{12}^j,\ \ \ S_{12}S_{34}-P_{12}P_{34},\ \ \
S_{12}P_{34}-P_{12}S_{34}.}
Therefore, if we restrict our attention to
those kinds of terms, again with an arbitrary structure of momenta, the
symmetry will be fulfilled.

Note that
the first type of operator in \chir\
carries a space index $j$ and two fields,
whereas the other two
have no indices and four fields. {}From this it can be immediately inferred
that with an even number of derivatives one can only have operators
with $4n$ fields ($n$
integer), whereas with an odd number of derivatives the allowed
operators contain $4n+2$ fields.  The reason is that all indices must be
contracted, either by the Kronecker delta $\delta^{ij}$ or by the complete
antisymmetric tensor in two dimensions $\epsilon^{ij}$, which
implies that an odd number
of derivatives needs an odd number of operators of the type $V^j_{12}$.

v) $\underline{Parity}$.
It only remains to impose discrete symmetries.  Parity is easy:  for
a Euclidean invariant operator, products of $S$, $V^j$ and momenta
are parity-conserving.  The only problem is when we have the
pseudo-scalar operator $P$.  What we have to do is just follow the
standard rule:  a term with an odd number of $P$'s must contain a
Levi-Civita symbol $\epsilon^{ij}$ also; a term with an even number of
$P$'s must not.

vi) $\underline{Charge\ conjugation}$.
To impose charge conjugation and reflection hermiticity
proves to be the most involved task.  This is because both operations
exchange fermions and antifermions, and thus they change, in general, the
momentum structure.  Explicitly, under charge conjugation our elementary
operators transform as
\eqn\chcon{S_{12}\rightarrow S_{21},\ \ \
P_{12}\rightarrow-P_{21},\ \ \ V^j_{12}\rightarrow-V^j_{21}.}
To take into account this symmetry at the level of the basis, the most
effective manner is to
consider all momenta written in combinations like
$\left(p_1\pm p_2\right)^j$, where $p_1$ is the momentum of an
antifermion and $p_2$ the momentum of the fermion of the same bilinear.
In this way it is easy to distinguish between C-conserving and
C-violating operators, and to construct both sets.

vii) $\underline{Reflection\ hermiticity}$.
The last one is reflection hermiticity.  It is defined, in principle, in
coordinate space \book\ and under such transformation, our
``elementary operators" behave just as in \chcon.  What is new is that
when transformed, one must change the coefficient of the operator by its
complex conjugate.  Therefore, once we restrict ourselves to
C-conserving terms, this additional symmetry restricts the coefficients
of those terms to be real.  The only subtlety is that, as it is defined,
the fields do not become complex conjugated and neither do their 
derivatives\foot{
We remind the reader that, in order to turn properly
from Minkowski to Euclidean space, one
has to redefine the symmetries of the problem, specially those which involve
complex conjugation.  Our definitions coincide with those of Ref.~\book.}.
And if one remembers that a derivative in coordinate space amounts to a factor
$-ip$ this indicates that an extra power of $i$ should be added for
each power of
momentum.

viii) $\underline{Further\ degeneracies}$.
Finally, the freedom of integrating by parts
(each operator has a delta function of momenta
conservation) relates different functionals, and, ultimately,
reduce the number of independent ones.
The best way of implementing these final
constraints is to find out a criterion in order to write down
every operator in a ``standard'' way.
We will explain ours in Appendix A, where we will also write down the
complete action, consisting of a basis of 107 functionals.

One of them is rather peculiar.  It is
\eqn\threeder{iV_{12}^j(p_1-p_2)^j(p_1-p_2)^2.}
One may be worried about it because it would
lead to a propagator with an additional pole besides the physical one on
the particle mass-shell, thus entering in conflict with unitarity.
This is, however, not important at all, because the above kind
of reasoning implies that one assumes a well-defined perturbative
expansion, and this is not the case (we have irrelevant operators
that make any perturbative expansion around the Gaussian fixed
point completely ill-defined).  One should think that the theory is such
that it manages to have a well-defined complete
two-point function free from unphysical
singularities. A completely different point is that, besides the above
discussion, when one computes the $\beta$-functions of
the theory one realizes that this operator, at least up to the order we
are considering, does not contribute to any other.  Therefore its evolution
will affect absolutely no conclusion we obtain without it.  For this
reason, we do not include it in the action.  We should remark, however,
that for all we do it is as if this term were already there, although for the
sake of brevity we will not write it down any more.

\newsec{Computing the $\beta $-functions}

Once we have constructed the initial action we want to work with, the
next step is to substitute it into the ERG equation \feq\ and to compute
the $\beta$-functions of our model within the given approximation.

In principle this is just a purely algebraic exercise.  Nevertheless, it
turns out that from a practical point of view it is an almost forbidding
task, if done by hand.  During intermediate steps of the calculation one
has to handle thousands of terms and it is too easy to make errors.  For
instance, when computing ${\delta S\over \delta \psi_p}\dot P_{\Lambda
}\left( p\right){\delta S\over \delta \psibar_{-p}}$, the functional
differentiation gives 302 terms, and one has, roughly speaking, to
square them and multiply the result by the inverse propagator.  Then,
one has to compute the appropriate products of gamma matrices, expand
all the terms and, finally, perform the integration by parts to reach
our chosen basis.  The number of operators considerably increases in
these last processes.  Thus, it is mandatory to use a symbolic
manipulation computer program to perform the functional differentiation,
do the algebra and integrate by parts.  Because of this, our computation
was done with the help of {\it Mathematica}.

To calculate the flow equations, we use an extended action, greater than
that discussed so far, in order to have some extra check of our
equations.  That is, we consider an action expanded in a basis that
consists of terms with two fermions with one and three derivatives, four
fermions with zero and two derivatives and six fermions with one and
three derivatives, but without imposing any symmetry other than vector
$U(N)$, parity and Euclidean invariance (that is, we impose neither
reflection hermiticity, nor charge conjugation, nor chirality).  We then
project the space generated by this basis into the invariant subspace
under the required symmetries and its direct complement.  The required
flow equations are obtained after the first projection, while the
complementary subspace provide us a consistency check of the
calculation.  They define a set of null equations that have to be
satisfied along the renormalization flow:  after projecting to an
initial symmetric action, any non-zero contribution of a non-symmetric
term will indicate an anomaly, which we have argued are non-existing.
We leave to Appendix B the complete set of $\beta$-functions.

Finally let us justify the inclusion of operators with three derivatives
into the action.  As we advanced above, it is motivated by the fact that
those terms are necessary in order to get a non-vanishing critical
anomalous dimension.  The argument is as follows.  The anomalous
dimension is related to the fact that we are free to fix the
normalization of one term of the action, by choosing an appropriate
normalization of our fields.  If, as is customary, we keep fixed the
coefficient of the so-called kinetic term, then its $\beta $-function is
substituted by an equation for $\eta $ which, in practice, is calculated
in a similar fashion.  We have to study, then, which type of Kadanoff
transformations contribute to $\psibar_{-p}\rlap /p \psi_p$.  The term
${\delta S\over \delta \psi_p}i \rlap/p{\delta S\over \delta
\psibar_{-p}}$ cannot, unless there were a mass operator, which is
forbidden by chiral symmetry.  There are, however, some contributions
coming from ${\cal S}^{(4,2)}$, due to ${\delta \over \delta \psi_p}i
\rlap/p{\delta S\over \delta \psibar_{-p}}$.  We will find, hence, the
anomalous dimension as a linear combination of couplings of ${\cal
S}^{(4,2)}$ and, consequently, if these couplings vanish at the fixed
point then $\eta =0$ is unavoidable.  (We define ${\cal S}^{(a,b)}$ as
the part of the action that contains $a$ fields and $b$ derivatives).
If one now studies $\dot{\cal S}^{(4,2)}$, it is not difficult to
convince oneself that its only contributions must come from ${\cal
S}^{(6,3)}$, apart from canonical rescalings.  The implications are now
immediate:  if ${\cal S}^{(6,3)}$ did not exist, then the whole action
${\cal S}^{(4,2)}$ would evolve canonically, thus it would vanish at the
fixed point and we would obtain a vanishing anomalous dimension.

\newsec{Fixed points, critical exponents}

\subsec{Generalities}

The next step is to find the fixed point solutions, that is, the sets of
coupling constants that make all the $\beta$-functions vanish.  These
will indicate the points to which the RG tends to, thus providing us
with the first indication of what the phase diagram of the system looks
like.

The condition $\dot S=0$ is equivalent to a system of 106 non-linear
algebraic equations.  To simplify it we note that all the coupling
constants of operators with six fields must enter linearly, because the
only source of non-linearity of Eq.~\feq\ is its first term on the
r.h.s., and it can give contributions neither from ${\cal S}^{(6,1)}$
nor from ${\cal S}^{(6,3)}$, within our approximation.  Therefore, we
can reduce the system to a set of only 13 non-linear equations,
\eqnn\redeq
$$\eqalignno{
0  &=  2 \eta g_1
  +  8 g_1^2 \alpha \gamma  N/(-2 + 3 \eta)
    + \{8 g_1^2 \beta \delta  N
  +  8 \beta \gamma [
      g_1 ( - 4 r_2 + s_1 -  s_2 - 3 s_3 - 3 s_4 )
     \cr & + 2 g_2 ( m_1 - m_2 - m_3 )
       + 4 g_1 m_3  N]\}/(-4 + 3 \eta),
\cr
0  &=  2 \eta g_2 +
   8 g_1^2 \alpha \gamma/(-2 + 3 \eta) +
     \{ 8 g_1^2 \beta \delta +
   8 \beta \gamma [
     2 g_1 ( m_1 - m_2 + m_3 - s_2 )
     \cr & + g_2 ( - 4 r_2 + s_1
	- s_2 - 3 s_3 - 3 s_4 ) + 2 g_2 s_2  N ] \}/(
    -4 + 3 \eta),\cr
0  &=  2 (-1 + \eta) m_1 + \{ 16 g_1 \alpha \delta (-2 g_2
 + g_1  N) +
   4 \alpha \gamma [
     g_1 ( m_1 - m_2 - m_3
	- 2 r_1 - 6 r_3
     - 4 s_2 - 3 s_3 + s_4 - t)
      \cr & - g_2 ( 3 m_1 + m_2 + 5 m_3 + 2 t )
     + 2 g_1 N ( 2 m_1 +
     3 m_3 + r_2 + 2 s_3 +
	s_4 + t ) ] \}/(-4 + 3 \eta),\cr
0  &=  2 (-1 + \eta) m_2 + \{16 g_1 g_2 \alpha \delta
   + 2 \alpha \gamma [
    g_1 ( 2 r_1 + 2 r_2 + 6 r_3 - 3 s_1 + 5 s_2 + s_3 + 5 s_4 )
    \cr & + 2 g_2 ( m_1 + m_2 + 3 m_3 - t )
    ]\}/(-4 + 3 \eta),\cr
0  &= 2 (-1 + \eta) m_3 + \{16 g_1 g_2 \alpha \delta +
   2 \alpha \gamma [ g_1 ( 2 m_1 - 2 m_2 - 2 m_3
	+ 2 r_1 - 2 r_2 + 6 r_3
       - s_1 - s_2 - 7 s_3 - 3 s_4 - 2 t )
     \cr & + 2 g_2 ( 2 m_3 + 2 m_1 - t )
     + 4 g_1 N ( m_3  + r_2 +
	2 s_3 + s_4 ) ] \}/(-4 + 3 \eta),\cr
0  &=  2 (-1 + \eta) r_1 + \{
    8 g_2 \alpha \delta (2 g_1 - g_2  N) +
   4 \alpha \gamma [
      g_1 ( m_1 - m_2 + m_3 + 4 r_3 - 2 s_2 ) \cr &
     + g_2 ( 3 m_1 + m_2 + m_3 - 2 r_2
     + s_1 - 3 s_2 - 4 s_3 - 4 s_4 - t )
       + 2 g_2 N ( r_2  - 2 r_3 + s_2 +
	2 s_3 + s_4 ) ]\}/(-4 + 3 \eta),\cr
0  &= 2 (-1 + \eta) r_2 + \{4 (g_1^2 + 4 g_2^2) \alpha \delta +
   4 \alpha \gamma [ g_1 ( 2 m_3 - t ) + g_2 ( 2 r_1 + 6 r_3 - s_1
     + 3 s_2 ) ]\}/(-4 + 3 \eta),\cr
 0  &= 2 (-1 + \eta) r_3 + \{12 g_1^2 \alpha \delta +
   4 \alpha \gamma [ g_1 ( 3 m_1 + m_2 + 5 m_3 - t ) \cr &
    + g_2 ( m_1 - m_2 - m_3 - s_1 - s_2 - 3 s_3 + s_4 - t ) +
	2 g_2 N ( r_2  + 2 s_3 + s_4 ) ] \}/
    (-4 + 3 \eta),\cr
 0  &=  2 (-1 + \eta) s_1
 + \{16 g_2 \alpha \delta (2 g_1 - g_2  N) +
   16 \alpha \gamma [
     g_1 ( 2 r_3 + s_2 ) + g_2 ( m_1 + m_2 + m_3 ) \cr & -  g_2 N (
   2 r_3 + s_2 ) ] \}/(-4 + 3 \eta),\cr
 0  &=  2 (-1 + \eta) s_2 - \{8 g_1^2 \alpha \delta +
   8 g_1 \alpha \gamma (m_1 + m_2 + m_3)\}/(- 4 + 3 \eta),\cr
 0  &= 2 (-1 + \eta) s_3 + \{4 g_1^2 \alpha \delta +
   4 \alpha \gamma [ g_1 ( 2 m_1 + 2 m_2 + t ) + g_2 ( - 2 r_1 + 2 r_3 + s_1
     + s_2 )]\}/(-4 + 3 \eta),\cr
 0  &= 2 (-1 + \eta) s_4 + \{4 g_1^2 \alpha \delta +
   4 \alpha \gamma [ g_1 ( 2 m_3 - t ) + g_2 ( 2 r_1 - 2 r_3 - s_1 - s_2 )
     ]\}/(-4 + 3 \eta),\cr
 0  &=  2 (-1 + \eta) t + 8 \alpha \gamma [
     g_1 ( - 2 r_1 + 2 r_3 + s_1 + s_2 )
      + g_2 ( m_1 + 3 m_2 - m_3 )
     + g_1 N ( - 2 m_2 + t ) ] /(-4 + 3 \eta),&\redeq \cr
}$$
where $\eta$ is the anomalous dimension that turns out to be
\eqn\eqeta{
\eta = 4 \alpha [-m_1 + m_2 + m_3 + s_1 + s_2 + s_3 + s_4 + t - 2 N ( r_2 +
     2 s_3 + s_4 )].}
$N$ is the number of flavours
and $\alpha $, $\beta $, $\gamma $, $\delta $ are
scheme dependent parameters defined as
\eqn\spar{\alpha ={1\over (2\pi)^d}\int_pK'\left( p^2\right) ,
\ \ \ \beta ={1\over (2\pi)^d}\int_p p^2 K'\left( p^2\right) ,
\ \ \ \gamma =K'\left( 0\right) ,\ \ \ \delta =K''\left( 0\right)
.}

The appearance of the above quantities just reflects the freedom in
choosing a renormalization scheme.
Furthermore, although
the $\beta $-functions depend on four parameters, we will see that
the fixed point solution will depend only on two combinations of them.  This
is just the pattern that occurs in a scalar theory within a similar truncation
\BHLM.

The system cannot be solved analytically, unless we perform further
simplifications, like keeping only the dominant term in an asymptotic
expansion at $N\to \infty$.  On the other hand, one can, of course,
simply try to study its solution numerically.  We will present both
approaches in turn.

After the fixed points are identified, the behaviour of the theory near
each of them is controlled by the critical exponents.  One of them is
fixed once we solve our set of equations:  it is the anomalous dimension
at the fixed point value.  The rest are found by linearizing the RG
transformations near the chosen fixed point.  That is, if $g_i$ is a
generic coupling constant, then its variation in the vicinity of a fixed
point $g_0$ is approximated by $\delta \dot g_i=\dot
g_i=R_{ij}|_{g_0}\delta g_j$, where $R_{ij}$ is the matrix ${\partial
\dot g_i\over \partial g_j}$.  The eigenvalues of $R_{ij}|_{g_0}$ can be
identified with critical exponents.  They can be thought of as the
anomalous dimension of the operators which drive the theory away from
the fixed point.

We can now no longer work with the reduced system of $13$ couplings, but
the full $105\times 105$ matrix is needed as we allow deviations from
the fixed point values of the six-fermion couplings.  Around the
Gaussian fixed point, for instance, the four-fermion and two-derivative
operators have the same degree of ``irrelevance" as the six-fermion and
one-derivative ones.

\nref\PMS{P.M. Stevenson, \prd{23}{81}{2916}}
Finally, let us turn again to the issue of scheme dependence.  We have
just said that, in general, the precise values of the coupling constants
at a fixed point are scheme dependent, thus reflecting that they are not
universal quantities.  Critical exponents, on the other hand, are
universal, hence they should be scheme independent.  Nonetheless, due to
the truncation, scheme dependencies will inevitably appear.  What we
will do is, as usual, to try to find a suitable scheme where the
dependence will not be that important.  To this end, we will apply to
the various solutions a principle of minimal sensitivity to discriminate
among different results \refs{\PMS, \BHLM}.

\subsec{$N\to \infty$}

We are now going to set up a large $N$ expansion for our model, with
which analytic results can be obtained.  Later on we will see that when
we study the general case by suitable numerical approximations, we will
recover our present results as a first term of the asymptotic series
around $N\to \infty $.

To define properly our approximation, we substitute each coupling
constant $g_i$ by $N^{z_i}g_i$ and study the limit $N\to \infty $
keeping $g_i $ fixed.  In principle, $z_i$ can be any real number, but
for simplicity we only consider integer values.  We then find the set
$\left\{ z_i\right\} $ that makes
all $\beta $-functions finite and, if possible, non-trivial.

With the above requirements, there are essentially two different manners
to define the $1/N$ expansion, which lead to different results.  We
label them by I and II, and discuss each in turn.

The Type I solution is obtained by considering $z_i=-1$, where $i$ runs
over every of the couplings that enter in Eq.~\redeq.  With this
definition, the anomalous dimension vanishes at leading order in $1/N$
and the system \redeq\ becomes
\eqnn\nequations
$$\eqalignno{
0&=-4 \alpha \gamma g_1^2 - 2 \beta \delta g_1^2 - 8 \beta \gamma g_1 m_3=
-4 g_2 s_2 \beta \gamma ,\cr
0&=- 2 m_1 - 4 \alpha \delta g_1^2 -
   2 \alpha \gamma g_1 (2 m_1 + 3 m_3 + r_2 + 2 s_3 + s_4 + t)
=  -2 m_2 =  -2 m_3 - 2 \alpha \gamma g_1 (m_3 + r_2 +
    2 s_3 + s_4),\cr
0&=- 2 r_1 + 2 \alpha \delta g_2^2+
   2 \alpha \gamma g_2 (-r_2 + 2 r_3 - s_2 - 2 s_3 - s_4)
=  -2 r_2 =  -2 r_3 - 2 \alpha \gamma g_2 (r_2 + 2 s_3 +
    s_4),\cr
0&=- 2 s_1 + 4 \alpha \delta g_2^2 + 4 \alpha \gamma g_2 (2 r_3 +
    s_2)
=  -2 s_2 =  -2 s_3 =  -2 s_4,\cr
0&=- 2 t + 2 \alpha \gamma g_1 (2 m_2 - t).&\nequations \cr
}$$
Its solution is
\eqnn\soli
$$\eqalignno{
g_1  & = -1/( \alpha  \gamma),
\ \ m_2  = 0,\ \ m_3  =  \delta/(4  \alpha  \gamma^2) +
   1/(2  \beta  \gamma),
\ \ r_1  =  \alpha \delta g_2^2 ,\ \ r_2  = r_3  = 0,\cr
\ \ s_1  & = 2  \alpha  \delta g_2^2 ,\ \ s_2  = s_3  = s_4 = 0,
\ \ t  = 5  \delta/( 4 \alpha  \gamma^2) - 6/(4  \beta
	\gamma)-  m_1.&\soli \cr}$$

We now choose $z_i=-2$ for all the six fermions coupling constants.
This is not the only solution since there are other rescalings
consistent with the reduced system \nequations.  For example, one can
assign the value $-1$ to some of the $z_i$'s and $-2$ to the other ones,
but it turns out that the results below do not depend on that.

The characteristic polynomial $P(\lambda )$ associated to the matrix of
linear deviations is exactly computable,
\eqn\cpol{\eqalign{
P(\lambda)=&\lambda^2\ (\lambda+2)^{12}\ (\lambda +4)^{83}\ (\lambda+6)\
 (\lambda^2+ 6 \lambda -8)\cr &\times \left(-\lambda^5 - 12 \lambda^4+
(8 w-44) \lambda^3 +(64 w-16) \lambda^2+(32 w +64)\lambda-(128 w +256)\right),
}}
where $w=\beta \delta /(\alpha \gamma)$ and $z=\delta /\gamma^2$.
We can read the critical exponents from $P(\lambda)$.  There are $100$
scheme independent eigenvalues, most of them
coinciding with the Gaussian values $0, -2, -4$ and $-6$.  The
non-trivial ones are $-3+\sqrt{17}=1.1231...$ and $-3-\sqrt{17}=-7.1231...
$, and
the roots of the polynomial
\eqn\lpol{Q(\lambda)=
-\lambda^5-12 \lambda^4+(8 w-44) \lambda^3+(64 w-16) \lambda^2+
(32w +64)\lambda-(128 w +256),
}
which are all $w$-dependent.
If $w<0$, that corresponds, for instance, to the exponential cut-off
function $K_{\Lambda}(p^2)=e^{-\kappa {p^2\over {\Lambda}^2}}$, the more
relevant critical exponent is $\lambda_1= 1.1231...$

Note that the fixed point solution depends freely on $g_2$ and $m_1$.
This is the expected result for the chiral Gross-Neveu model because the
$U(1)$ Thirring like excitations (which in our action are controlled by
$g_2$) decouple from the rest and this subsystem is conformally
invariant (i.e.~it is at fixed point) for any value of $g_2$.  For the
$SU(N)$ part there exists a discrete set of fixed points, the one of
Dashen and Frishman being one of them.  This critical point is reached
when the constant $g_1$ is of order $1/N$, as in our case.  So we can
make a correspondence between our solution and that of Ref.~\DF.
Nevertheless, the values of the anomalous dimension in both cases do not
match.  For the cited fixed point it is non-vanishing at leading order
in $1/ N$, and not zero as we have found.  This discrepancy with the
exact result of Dashen and Frishman could be caused by our truncation.
We cannot reject, however, the possibility of having found a different
fixed point as it has already occured previously \ofp.

For the Type II solution it is useful to define the new variable
$m_1'=m_1- m_2+m_3$ instead of $m_1$.  Then it corresponds to
the following rescaling of couplings
\eqnn\resci
$$\eqalignno{
g_1 &\to g_1/N,\ \ g_2  \to g_2/N,\ \
m_1' \to m_1',\ \
m_2  \to m_2/N,\ \ m_3  \to m_3/N,\ \
r_1  \to r_1/N,\ \
r_2  \to r_2/N,\ \ r_3  \to r_3/N,\cr
s_1  &\to s_1/N,\ \
s_2  \to s_2/N,\ \ s_3  \to s_3/N,\ \ s_4  \to s_4/N,\ \
t \to t.&\resci \cr
}$$
The solution takes the form
\eqnn\solii
$$\eqalignno{
g_1=&\beta {2 \alpha - \beta \gamma\over 48 \alpha^3},\ \
g_2=\beta {-2 \alpha + \beta \gamma\over 24 \alpha^3},\cr
m_1'= &{-8 \alpha + \beta \gamma\over 144 \alpha^2},\ \
m_2= {8 \alpha - \beta \gamma\over 72 \alpha^2},\ \
m_3=
{24 \alpha^2 \beta \delta + 32 \alpha^3 \gamma - 18 \alpha \beta^2
    \delta \gamma - 18 \alpha^2 \beta \gamma^2 + 3 \beta^3 \delta \gamma^2 +
     4 \alpha \beta^2 \gamma^3\over
576 \alpha^3 \gamma (-4 \alpha+
    \beta \gamma)},\cr
r_1= &{43\over 36 \alpha} + {\beta^2 \delta\over 12 \alpha^3} -
    {\beta \delta\over 6 \alpha^2 \gamma} - {43
\beta \gamma\over 288 \alpha^2},\ \
r_2= {-8 \alpha + \beta \gamma\over 288 \alpha^2},\ \
r_3= {8 \alpha -  \beta \gamma\over
    36 \alpha^2},\cr
s_1= &{20\over 9 \alpha} + {\beta^2 \delta\over 6 \alpha^3} -
    {\beta \delta\over 3 \alpha^2 \gamma} - {5 \beta \gamma\over
      18 \alpha^2},\ \
s_2= {8 \alpha - \beta \gamma\over 144 \alpha^2},\ \ s_3=
     {-8 \alpha + \beta \gamma\over 288 \alpha^2},\ \
s_4= {-8 \alpha + \beta \gamma\over 288 \alpha^2},\cr
t= &{8 \alpha - \beta \gamma\over 144 \alpha^2},&\solii \cr
}$$
which has a non-zero anomalous dimension,
\eqn\etaii{\eta={4\over 3} - {\beta \gamma\over 6\alpha}. }

The set of the remaining $z_i$'s is unique and composed of the numbers
$-1$ and $-2$.  Unfortunately, unlike the previous case we could not
find the exact analytical expression of the characteristic polynomial.
However by computing numerically the eigenvalues for different values of
$z$ and $w$ we could guess some exact results.  None of the critical
exponents coincide with their canonical counterparts.  Moreover, most of
them are functions of the combination ${w\over z}$.  Thus there are 82
eigenvalues $\lambda=-{w\over 2z}$, 8 eigenvalues equal to ${2\over
3}(1-{w\over 2z})$ and 4 of the form $2-{w\over 2z}$.  The remaining
ones are not functions of the ratio $w/z$ only (and even a few have a
non-vanishing imaginary part, which is not unusual in approximations
based on truncations).  We have to study numerically the most relevant
critical exponent, which belongs to the class with no simple dependence
in $w$ and $z$, for different scheme parametrizations.  As it happens in
the scalar case, for any value of $z$, this exponent always has a
minimum at some scheme parametrized by $w=w^*$.  This behaviour induced
us to use the minimal sensitivity criterion to fix the parameter $w$ to
its critical value $w^*$.  Unfortunately due to the monotonic
dependence of the solution on the parameter $z$ in the range analyzed,
we were unable to set it with a similar prescription.  In Table I we
show some values of $\lambda_1^*=\lambda_1(w^*)$ and the anomalous
dimension for various $z$'s.

\topinsert\hfil
$$\vbox{\tabskip=0pt \offinterlineskip

\halign to 300pt{\strut#& \vrule#\tabskip=1em plus2em&
   \hfil#& \vrule#&
   \hfil#& \vrule#&
   \hfil#& \vrule#& \hfil#& \vrule#&
   \hfil#& \vrule#\tabskip=0pt\cr
\noalign{\hrule}
& &\hidewidth \hidewidth
& &\hidewidth $z=0.1$ \hidewidth
& &\hidewidth $z=0.5$ \hidewidth
& &\hidewidth $z=1.0$ \hidewidth
& &\hidewidth $z=2.0$ \hidewidth &
 \cr \noalign{\hrule}
 &&{$\lambda^*_1$}  &&$2.258$  &&$2.239$  &&$2.217$  &&$2.175$
						      & \cr\noalign{\hrule}
 &&{$w^*$}  &&$0.122$  &&$0.616$  &&$1.250$  &&$2.610$
						      & \cr\noalign{\hrule}
 &&{$\eta$}  &&$1.130$  &&$1.128$  &&$1.125$  &&$1.116$
						      & \cr\noalign{\hrule}
\noalign{\smallskip}}}$$
\hfil
\centerline{\vbox{\hsize= 350pt \noindent\footnotefont
Table~I: Local minimum of $\lambda_1$, the most relevant critical exponent,
corresponding to the solution for $N\to \infty$
labelled as Type II in the text,
for different values of $z$.
$\eta$ is the anomalous dimension at that point and $w^*$ is the value
of the parameter $w$ at which the minimum is
reached.}}
\bigskip
\endinsert

\subsec{Finite $N$}

For a finite number of flavours analytical results for the fixed point
couplings cannot be found.  So one has to proceed numerically to search
for the zeroes of the $\beta$-functions.  Moreover, the number of
different solutions of a system of coupled non-linear equations is not
known a priori and the common routines for root-finding (such as the
FindRoot command of {\it Mathematica}) do not guarantee that all the
zeroes are reached.  A more serious inconvenience is to decide if a
given zero corresponds to a real fixed point solution or if it is a
spurious root resulting from our truncation.

The first problem can be reasonably reduced after some experience is
acquired.  In fact, we can know by intuition which is a reasonable range
of values for the couplings and inspect this region exhaustively.  Of
course this is not easy for a system of thirteen equations, but we can
gain some confidence in the results if we examine minutely the adequate
region.

The second problem, however, is much more complicated.  In principle we
do not have any criteria to decide if a root of the $\beta$-functions
system corresponds to a genuine fixed point solution or if it is a
fictitious artifact of our approximations.  This problem, which already
appeared in the bosonic case too, is perhaps the Achilles' heel of the
approximations based on truncations \refs{\spura, \spurb}.  We present
the class of solutions of which we are more confident.  These are mainly
the ones which asymptotically match with some solution clearly
identified in the framework of the large $N$ expansion.

Let us, to begin with, select a particular scheme and find the solution
for different values of $N$:  $w=-2$ and $z=0.5$, corresponding to an
exponential regulating function ($K(x)=e^{-x^2}$).  We analyze the
dependence on these parameters later on.

One solution is found which behaves asymptotically as the type I one in
the $N \to \infty$ limit.  $N \eta$ increases with $N$ and tends to
$4.87...$, while the most relevant critical exponent $\lambda_1$
decreases with $N$ asymptotically to the value $1.1231...$, in agreement
with the $1/N$ expansion.  For the second eigenvalue we find complex
figures that we attribute to our approximations.  Another piece of bad
news is that, unlike the $N \to \infty$ case, the solution for finite
though big enough $N$, is isolated, while, as we mentioned before, the
fixed point solutions for Thirring like models are continuous in the
$U(1)$ sector.  Again we blame this confusing result on the truncation.
We present in \fig\neta{$N\eta$ (solid line) and $\lambda_1$ (dashed
line) as functions of $N$.  This solution matches with the Type I
solution of the large $N$ limit.} the curves for $N \eta$ and
$\lambda_1$ as a function of $N$.

More interesting is perhaps the study of the dependence of the solution
on the scheme.  We have noticed that $z$ enters the equations only
through the anomalous dimension as a global factor.  For this reason,
the dependence of the fixed points solutions in $z$ is quite simple:  it
is almost linear in $\eta$.  Therefore it is more attractive to
investigate its behaviour under changes on the parameter $w$ for fixed
$N$ and $z$.  The motivation of this analysis is the search, as in the
scalar case, of some non-linear $w$-dependence in such a way that we can
invoke a principle of minimum sensitivity to fix the value of this
parameter and eliminate one fictitious dependence.  To this end, we fix
the value of $z$ to $z=0.5$ and $N$ to $N=1000$.  The curve $\eta$
vs.~$w$ is monotonic and decreases with $w$, while the first eigenvalue
$\lambda_1$ reaches its minimum value $\lambda_1=1.12511$ at $w=-45$,
which increases as we lower $N$:  it is equal to $1.1273$ for $N=500$
(it is reached at $w=-23$), $1.146$ for $N=200$ (at $w=-10$), $1.1519$
for $N=100$ (at $w=-8$), $1.695$ for $N=10$ (at $w=-2.4$) and finally,
$2.560$ for $N=3$ (at $w=-0.5$).  We show two of these curves in
\fig\dn{$\lambda_1$ as a function of $w$, ($z=0.5$), for $N=3$ and
$N=10$.  The minimum clearly decreases with $N$.}.

For the fixed point that matches the Type II solution as $N\to \infty$
we found a curious behaviour.  For $N$ moderately large, (say $N=1000$),
the numerical solution is in good agreement with the $1/N$ analytical
result (for example, the value of $\eta$ for $z=0.5$ and $w=-2$ is
$1.99$, compared with the exact $\eta=2$ for $N\to \infty$).  As we
lower $N$ the values of the anomalous dimension $\eta$ and the most
relevant eigenvalue $\lambda_1$ decrease.  But, unexpectedly, the
solution disappears at $N=142$ (actually at $N=142.8$ if we let $N$ take
non-integer values).  A closer analysis of the space of solutions shows
us that at this value of $N$ the branch of solutions compatible with the
type II $1/N$ expansion merges with another family of fixed points.
This last branch has finite asymptotic limits for $\eta$ and $\lambda_1$
as $N\to \infty$.  However some couplings do not behave as a power of
$N$ in this limit and, therefore, it cannot be associated with a $1/N$
fixed point in the sense stated previously.  At the bifurcation point
$\eta= 1.88$ and $\lambda_1= 5.80$.  We show in \fig\netaii{$\eta$
(solid line) and $\lambda_1$ (dashed line) as a function of $N$ for
$z=0.5$ and $w=-2$.  In both curves the upper branch corresponds to the
solution that matches with the Type II large $N$ solution.} the curves
$\eta(N)$ and $\lambda_1(N)$.  This peculiar behaviour of the type II
solution suggests that it cannot be identified with the Dashen and
Frishman fixed point, which exists for any value of $N$.  Even though
this behaviour could be another consequence of the truncation, it is
hard to justify it because for low $N$ only operators with few spinor
fields are allowed by the Pauli principle, and thus we expect our
six-fermions truncation to be accurate.  We have not been able to solve
this puzzle.

It is easy to find many other solutions, especially for low $N$.  For
some of them there exists a minimum, either for $\eta$ or for
$\lambda_1$ but in other cases both curves are monotonic in $w$.  They
have also different behaviours as $N\to \infty$.  We show one example in
\fig\osol{$\eta$ (solid line) and $\lambda_1$ (dashed line) as a
function of $N$, ($z=0.5$, $w=-2$) for a different fixed point solution.
In this case both exponents are of order $1$ as $N\to \infty$.}.

\subsec{$N=1$}

Finally we will consider the special case $N=1$. For this particular
value of $N$ not all the operators presented in Appendix A
are independent.  As we mentioned in Sect.~3 the effect of the
Fierz transformations is to relate covariant $U(N)$ local operators (like
$\psibar^a\left(p_1\right)\psi^b\left(p_2\right)$) to scalar ones (like
$\psibar^a\left(p_1\right)\psi^a\left(p_2\right)$). So in the $N=1$ case
the Fierz transformations uncover relations between the $S$, $P$ and
the $V^j$ operators. For example, for operators without derivatives we have
the identities
\eqn\nunoiden{
S_{12}S_{34} = -P_{12}P_{34} = -{1\over 2} V_{12}^j V_{34}^j .}

They establish relations between the coupling constants that permit us
to reduce considerably the system.  For the set of couplings of the four
fermions operators, the independent ones are
\eqnn\indep
$$\eqalignno{
{\tilde g}&=g_1 - g_2,\ \ \ \ \ {\tilde u}_1=m_1 - m_2 + m_3 - r_1 +
 r_2 - r_3 + (s_1 - s_2 + s_3 + s_4)/2,\cr
{\tilde u}_2&= 2 m_1 + 2 m_2 - 2 m_3 + 2 r_1 + 2 r_2 - 2 r_3,\ \ \ \ \
{\tilde u}_3=  4 m_2 + 2 r_1 - 2 r_2 - 2 r_3 - 2 s_3 - 2 s_4,\cr
{\tilde u}_4&= -s_1 - s_2 + s_3 + s_4,\ \ \ \ \
{\tilde u}_5= -2 s_3 + 2 s_4 + 2 t.&\indep \cr}$$

Eq.~\redeq\
is now a 7-equation system
that looks like
\eqnn\fpnuno
$$\eqalignno{
0 &= 2 {\tilde g} (4 \eta - 3 \eta^2 + 4 {\tilde u}_2 w - 4 {\tilde u}_3
w + 4 {\tilde u}_4 w)/ (4 - 3 \eta) \cr
0 &= 2 (8 {\tilde g}^2 + 4 {\tilde u}_1 + 8 {\tilde g} {\tilde u}_1 + 2
{\tilde g} {\tilde u}_3 + 2 {\tilde g} {\tilde u}_4 - 7 {\tilde u}_1 \eta
+ 3 {\tilde u}_1 \eta^2)/(-4 +3 \eta)\cr
0 &= 2 (8 {\tilde g}^2 + 8 {\tilde g} {\tilde u}_1 + 4 {\tilde u}_2 + 8
{\tilde g} {\tilde u}_4 + 4 {\tilde g} {\tilde u}_5 - 7 {\tilde u}_2 \eta +
3 {\tilde u}_2 \eta^2)/(-4 + 3 \eta)\cr
0 &= 2 (-24 {\tilde g}^2 - 24 {\tilde g} {\tilde u}_1 + 4 {\tilde g}
{\tilde u}_2 + 4 {\tilde u}_3 - 12 {\tilde g} {\tilde u}_3 + 4 {\tilde
g} {\tilde u}_5 - 7 {\tilde u}_3 \eta + 3 {\tilde u}_3 \eta^2)/(-4 + 3
\eta) \cr
0 &= 2 (8 {\tilde g}^2 + 8 {\tilde g} {\tilde u}_1 + 4 {\tilde g}
{\tilde u}_3 + 4 {\tilde u}_4 + 4 {\tilde g} {\tilde u}_4 - 7 {\tilde u}_4
\eta + 3 {\tilde s} \eta^2)/(-4 + 3 \eta) \cr
0 &= 2 (-4 {\tilde g} {\tilde u}_2 - 4 {\tilde g} {\tilde u}_3 - 8
{\tilde g} {\tilde u}_4 + 4 {\tilde u}_5 - 7 {\tilde u}_5 \eta +
 3 {\tilde u}_5 \eta^2)/ (-4 + 3 \eta)\cr
\eta &= -2 {\tilde u}_2 z + 2 {\tilde u}_3 z - 4 {\tilde u}_4 z +
2 {\tilde u}_5 z.&\fpnuno \cr}$$
It is linear in
${\tilde u}_1$, ${\tilde u}_2$, ${\tilde u}_3$, ${\tilde u}_4$
and ${\tilde u}_5$, so we can solve it for these variables ending with a
two-equation system for ${\tilde g}$ and $\eta$.  After a bit of algebra and
discarding the trivial solution we finally get a unique equation for $\eta$,
\eqn\finsys{\eqalign{
0&=-120 w^2  z + 288 w z^2  + \eta (13 w^2  - 132 w z +
210 w^2  z + 288 z^2  - 720 w z^2 )\cr
&\ \ \ \ + \eta^2  (99 w z - 90 w^2  z - 432 z^2  + 594 w z ^2) +
 \eta^3  (162 z^2  - 162 w z^2 ) .
}}
As in the previous analysis we have to choose some particular scheme,
i.e.~fix $w$ and $z$, and solve the equation numerically.
Unfortunately, a simple inspection of the equation reveals bad news.
The system is not undetermined and there is no room for a free
${\tilde g}$-dependence of the fixed point solution as it is true in the
Thirring model.  This property is satisfied in the previous order
approximation (terms with less than three derivatives) where $\eta$
vanishes identically.  The reason this property is lost in the three
derivatives approximation is unclear for us.  For a more detailed
analysis it is necessary to go to the next order to see whether this
property is restored.

We solved Eq.~\finsys\ numerically for different values of $w$ and $z$.
As in previous examples the fixed point solutions are almost linear in
the parameter $z$ so it is more interesting to study the behaviour of
the solution as a function of $w$.  However, in the range of values
studied, we did not find any non-monotonoic behaviour either in the
critical couplings or in the anomalous dimension.  We present some of
the results in Table II.

\topinsert\hfil
$$\vbox{\tabskip=0pt \offinterlineskip

\halign to 300pt{\strut#& \vrule#\tabskip=1em plus2em&
   \hfil#& \vrule#&
   \hfil#& \vrule#&
   \hfil#& \vrule#& \hfil#& \vrule#&
   \hfil#& \vrule#\tabskip=0pt\cr
\noalign{\hrule}
& &\hidewidth \hidewidth
& &\hidewidth $w=-0.1$ \hidewidth
& &\hidewidth $w=-0.5$ \hidewidth
& &\hidewidth $w=-1.0$ \hidewidth
& &\hidewidth $w=-2.0$ \hidewidth &
 \cr \noalign{\hrule}
 &&{$z=0.1$}  &&$1.763$  &&$3.691$  &&$6.316$  &&$11.747$
						      & \cr\noalign{\hrule}
 &&{$z=0.5$}  &&$1.418$  &&$1.790$  &&$1.418$  &&$3.388$
						      & \cr\noalign{\hrule}
 &&{$z=1.0$}  &&$1.376$  &&$1.559$  &&$1.811$  &&$2.349$
						      & \cr\noalign{\hrule}
 &&{$z=2.0$}  &&$1.354$  &&$1.445$  &&$1.569$  &&$1.834$
						      &\cr\noalign{\hrule}
\noalign{\smallskip}}}$$
\hfil
\centerline{\vbox{\hsize= 350pt \noindent\footnotefont
Table~II: Values of $\eta$ for $N=1$ and different scheme parameters
$z$ and $w$.}} \bigskip
\endinsert

\newsec{Summary and discussions}

In this article, we analyze the application of the ERG method to
fermionic theories.  An ERG equation for Grassmann variables is derived
and the critical properties of the chiral Gross-Neveu model in two
dimensions are studied with it.

To solve the ERG equation, a non-linear functional differential
equation, we perform a double truncation, in the number of derivatives
(derivative expansion) and in the number of fields (polynomial
approximation).  Unfortunately, these approximations produce similar
problems that already appear in the scalar case within analogous
truncations:  spurious solutions and unphysical scheme dependencies.
The latter, which is a common feature of almost any approximation in
QFT, can be partially disentangled by invoking a minimum sensitivity
criterion:  for a given observable we choose the scheme that gives the
most ``stable'' result.  The emergence of spurious solutions is a more
serious problem.  In principle we do not have any strong argument to
accept or reject a solution, except for those which lead to absurd
results.

Note that for the bosonic case, within the derivative expansion, we can
either expand the action as a polynomial in the fields, leading to a
system of coupled non-linear equations or not make any further
approximation and consider the potentials as arbitrary functions (not
necessarily real analytical), that requires the study of partial
differential equations.  While the first approach produces lots of
spurious solutions of the fixed point equations \refs{\spura, \spurb},
the former has shown to produce the correct ones \refs{\der, \BHLM}.
For the fermionic case, however, the situation is quite different.  For
finite $N$, within the derivative expansion, a truncation in the number
of fields is not an approximation for local Lagrangians, but the
definition of a function in terms of Grassmann variables.  So, for
fermions, the polynomial approximation should not produce fictitious
solutions if we are constraining the number of derivatives.  In
accordance with this, for $N=1$, which is the only case where we
actually have a pure derivative expansion without any further
truncation, no spurious solution appears.  (There are only three
solutions, besides the trivial one, and two of them have complex
coupling constants, thus being rejected at once).  It would be
interesting to perform a pure derivative expansion for, say, $N=2$ and
check if the above pattern holds.

The first analysis we do of the fixed point structure of the chiral
Gross-Neveu model is in the large $N$ expansion of the
$\beta$-functions.  We find that it can be defined in two different
ways, with remarkably different results.  The first one leads to a
continuous family of fixed-points which reminds that of Dashen and
Frishman:  the solution is free in the direction associated to the
abelian degrees of freedom and fixed of order $1/N$ in the direction of
the $SU(N)$ ones.  However it presents an important difference:  the
anomalous dimension vanishes at leading order in contrast to the order 1
value of the Dashen and Frishman solution.  We attribute this difference
to the truncation.  The inclusion of more terms should clarify this
point.

The other type of solution is much more involved.  Its anomalous
dimension is non-zero but the Thirring like excitations do not appear.
Moreover, unlike the preceding case its dominant eigenvalue of the
linearized RG transformation depends on the scheme.  Another astonishing
result is the structure of the remaining eigenvalues.  One would expect
that the most irrelevant ones would not be too different from the
canonical ones due to our truncation, which is not the case.
Furthermore, the stability of the solution for finite $N$ seems to
indicate that it is not an artifact of the truncation, but a true fixed
point.

To go further we proceed numerically.  We can clearly identify a
solution that asymptotically matches the first of the above ones and
trace it to very low values of $N$.  It presents two important
drawbacks.  The first one is that it is isolated, unlike the strict
limit $N\to \infty $, where a one-parameter space of solutions appears.
The other one is again a remaining dependence on a parameter which
labels different schemes, although an accurate analysis of the most
relevant critical exponent exhibits minimum sensitivity to some schemes.
A search of the behaviour of the critical exponents as a function of $N$
clearly shows that the value of $\lambda_1$ decreases with $N$ while
$N \eta$ increases.

We can also find another set of solutions, for $N>142$ that matches the
second type of the large $N$ ones.  At $N=142.8$ it merges with another
family of fixed points, with divergent $\beta $-functions when $N\to
\infty$, although its critical exponents seem to be finite in that
limit.  This odd behaviour ruled out an identification of this solution
with that of Dashen and Frishman.  The lack of information in the
literature (anomalous dimensions, critical exponents) about the extra
fixed points of the model does not allow us to recognize our solution as
any of them.

Finally, due to its peculiarities, we separately analyze the $N=1$ case.
The results are, however, discouraging.  On one hand, at first order in
the derivative expansion, there appears a solution with a free
parameter, which labels the Thirring like excitations, but it gets
spoiled when higher orders are considered.  On the other hand, we find a
severe two-parameter scheme dependence, which made unreliable any
conclusion.

Let us remark that the insufficient non-perturbative studies (lattice
computation, etc.) of the chiral Gross-Neveu model in $d=2$ impede to
discriminate definitively in favour of our results.  However we are very
confident of some of our findings:  as we argue above, within the $1/N$
expansion the first fixed point solution is an excellent candidate for
the Dashen-Frishman fixed point, whereas the other one presents
evidences to be a new fixed point, with quite intricate properties, not
discussed previously in the literature.  Moreover, both solutions have a
smooth behaviour for finite $N$.

\nref\redund{T.L. Bell
and K.G. Wilson, \prb{10}{74}{3935}; {\bf 11} (1975) 3431\semi
K.E. Newman and E.K. Riedel, \prb{30}{84}{6615}\semi
E.K. Riedel and G.R. Golner and K.E. Newman, \aps{161}{85}{178}}
We want to pause here to make a comment about redundant operators, that is,
those operators that do not affect correlation functions
\wegner.  Examples are
functionals proportional to the equations of motion.  They are known to
be present in general and, specifically, one should expect the existence of a
redundant operator that reflects the freedom of changing the normalization of
our action (e.g. modifying the coefficient of the kinetic term).  Its
associate eigenvalue is scheme dependent and of no physical relevance.
However,
the ERG equation we consider presents an invariance under the
normalization of the fields \intj\ which is not preserved by our
approximation (not even in a pure derivative expansion).
The signal of its restoration must be an appropriate redundant operator with
vanishing eigenvalue.  Its appearance should, fortunately, fix the anomalous
dimension close to its correct value.
It should also be mentioned that the presence of this operator has been used
several times in the literature to discriminate among different schemes
and find the appropriate anomalous dimension
\refs{\redund, \redtwo}.
We expect that similar techniques should apply for fermions also, although
we have not gone through the details.  More work has to be done in this
direction.

As a summary, we may say that, in general, our results seem somewhat
discouraging, specially those for low $N$.  Let us note, however, that
the two dimensional world is rather peculiar, due to the importance of
quantum effects, which generally produce large anomalous dimensions.
Thus, technical simplicity turns into increasing complexity while the
dimension is lowered.  Nevertheless, we have gone much further than
similar computations for bosonic theories, where in $d=2$ it seems that
the method completely breaks down \spura.  Another interesting feature
is the seemingly good results for the large $N$ limit.  At the
computational level this is related to the fact that at the leading
order of the $1/N$ expansion the system \redeq\ of the flow equations
simplifies dramatically and no room is left for spurious solutions.
Actually, improvement of results in the large $N$ limit appears to be a
general feature of the ERG approach (see, for instance, Ref.~\weq).
Therefore, the credibility of the method can only be clearly decided
after extensions to other dimensions and, possibly, the inclusion of
higher order terms.

A last comment is dedicated to further work.  As we have just mentioned,
the formalism should be extended to higher dimensions.
Equation \feq\ is prepared for that.  What has to be done is to
choose an appropriate action and perform a similar calculation.  The
number of spinor structures will be increased and, therefore, one will
have to handle more terms in the action.  However, we guess that,
due to the greater complexity, two derivatives may be sufficient to
obtain interesting results, or at least, according to the standard rule
that quantum effects become less important when the dimension is
increased, we hope that, within the same approximation, the results will
be more transparent.

\bigbreak\bigskip\bigskip\centerline{{\bf Acknowledgements}}\nobreak

We are sincerely indebted to J.I.~Latorre, who suggested to begin this
work and provided us fruitful suggestions during its completion.  We
acknowledge G.R.~Golner for interesting comments about the presence of
redundant operators.  Thanks are given to P.E.~Haagensen for reading the
manuscript and M.~Bonini for informing us about the work of Ref.~\QED.
Discussions with A.A.~Andrianov and D.~Espriu are also acknowledged.
J.C. and Yu.\ K.
thank T.R.~Morris for discussions about several aspects of this
and related subjects.  Yu.~K.~and E.~M.~would like to thank the
Department ECM of the University of Barcelona for warm hospitality and
friendly atmosphere during their stay there.

\appendix{A}{Conventions and notation}

In this appendix we summarize our notational conventions.

We are working always on Euclidean momentum space.  A standard term in the
action should be
\eqn\notact{\int_{p_1}\int_{p'_1}\dots\int_{p_n}\int_{p'_n}
{1\over (2 \pi)^{2 n d}}
\delta_{p_1+p'_1+\dots+p_n+p'_n}\psibar^{a_1}_{p_1}
\Gamma_1\psi^{b_1}_{p'_1}\dots\psibar^{a_n}_{p_n}\Gamma_n\psi_{p'_n}^{b_n}
\times ({\rm polyn.~in~momenta})\times({\rm flavour~matrices}),}
where
\eqn\momdel{\int_p\equiv\int d^dp\ ,
\ \ \ \ \ \ \delta_{p_1+p'_1+\dots+p_n+p'_n}
\equiv {(2 \pi)}^d\delta^d\left(\sum_{i=1}^n(p_i+p'_i)\right)\ ,}
the fermions are denoted
indicating its momentum label as a subindex, its flavour label as a superindex
(and usually using Latin letters of the begining of the alphabet) and the
``Clifford" label supressed; $\Gamma$ is a matrix that acts on Dirac indices
(if it carries a Lorentz index we denote it with a Latin index of the middle of
the alphabet).  Powers of momenta (corresponding to derivatives in
position space) are always indicated explicitly.  For instance, the usual
kinetic term would be
\eqn\notkin{
{1\over (2\pi)^{2d}}
\int_p\int_{p'}\psibar^a_pi\gamma^j{p'}^j\psi^a_{p'}\delta_{p+p'}
={1\over (2\pi)^d}\int_p\psibar^a_{-p}\gamma^jp^j\psi^a_p}
We repeatedly use Fierz transformations in order to reduce bilinears
to the diagonal in flavour indices form,
\eqn\notbil{S_{12}\equiv
\psibar^a_{p_1}\psi^a_{p_2},\
\ \
P_{12}\equiv \psibar^a_{p_1}\gamma_s\psi^a_{p_2},\
\ \
V_{12}^j\equiv \psibar^a_{p_1}\gamma^j\psi^a_{p_2}.}
When we do so all flavour matrices disappear.

Even more, for the sake of simplicity, we often drop the integral signs,
the delta functions and the powers of $(2 \pi)^d$.  However, when a
term of the action is written they should always be understood to be
present.  Also, for the action in Appendix B it is convenient to define
the sum and difference of the momenta of every bilinear.  For instance,
the above kinetic term will be written as \eqn\notkinpm{{1\over
2}p_{12}^{-\,j}iV_{12}^j} with $p_{mn}^{\pm\,j}\equiv (p_m\pm p_n)^j$
and with the integrals and momentum conservation delta function being
assumed.

The conventions for the Clifford algebra are the usual ones on the
two-dimensional Euclidean space \book,
\eqn\notclif
{\left\{\gamma^i,\gamma^j\right\}=2\delta^{ij},\ \ \
\gamma_s=-i\gamma^1\gamma^2.} The completely antisymmetric tensor
$\epsilon^{ij}$ is $\epsilon^{12}=-\epsilon^{21}=1$.

We always use $N$ to indicate the number of flavours and $\eta$ to denote the
anomalous dimension of the fields.  The conventions for the coupling constants
are a Latin or Greek letter with a subindex.  The letters are: 1) $g$ for the
four-fermions, non-derivative operators,
\eqn\notfourf{g_1(S_{12}S_{34}-P_{12}P_{34})+g_2V_{12}^jV_{34}^j\,;}
2) $m$, $r$, $s$ and $t$ for the various types of
four-fermions, two-derivatives operator
couplings; 3) $a$, $c$ and $e$ for the three
types of six-fermions, one-derivative operator
couplings; 4) Greek indices $\ks$,
$\kv$, $\jv$, $\ay$ and $\by$ for the six-fermions and
three-derivatives operator couplings.

When we expand the full action we introduce the scheme-dependent parameters
\eqn\notsche{
\alpha ={1\over (2\pi)^d} \int_pK'\left( p^2\right) ,\ \ \ \beta =
{1\over (2\pi)^d}\int_p p^2 K'\left( p^2\right) ,\ \ \ \gamma =
K'\left( 0\right) ,\ \ \ \delta =K''\left( 0\right)
.}
Sometimes they enter solely in the combinations $\omega={\beta\delta\over
\alpha\gamma}$, $z={\delta\over\gamma^2}$.

A dot in any quantity means a derivative with respect to the RG flow parameter
$t$.

\appendix{B}{The action}

In this appendix we present the complete action we use for the
computation of the $\beta$-functions.  For the sake of clarity the
action is divided into subactions according to the number of fermions
and the number of derivatives, and in the case of six fermions and three
derivatives also according to the fermionic structure.  The integrals
over the momenta and the $\delta$-functions of global momentum
conservation (with their respective powers of $(2 \pi)^d$) are always
omitted. The notation is presented in the Appendix A.
\eqnn\sss
$$\eqalignno{
{\cal S}^{(2,1)}&={1\over 2}p_{12}^{-\,j}iV_{12}^j.\cr
{\cal S}^{(4,0)}&=g_1(S_{12}S_{34}-P_{12}P_{34})+g_2V_{12}^jV_{34}^j.\cr
{\cal S}^{(4,2)}&=\{ m_1p_{12}^{+\,2}+m_2p_{12}^-\cdot p_{34}^-
+m_3p_{12}^{-\,2}\}\times (S_{12}S_{34}-P_{12}P_{34})
+\{r_1p_{12}^{+\,2}+r_2p_{12}^-\cdot p_{34}^-+r_3p_{12}^{-\,2}\}\times
V_{12}^jV_{34}^j\cr
&+\{s_1p_{34}^{+\,j}p_{12}^{+\,k}+s_2p_{12}^{-\,j}p_{12}^{-\,k}
+s_3p_{12}^{-\,j}p_{34}^{-\,k}+s_4p_{34}^{-\,j}p_{12}^{-\,k}\}\times V_{12}^j
V_{34}^k
+tp_{34}^{+\,j}p_{34}^{-\,k}\epsilon^{jk}(S_{12}P_{34}-P_{12}S_{34})
.\cr
{\cal S}^{(6,1)}&=\{a_1p_{12}^{-\,j}+a_2p_{56}^{-\,j}\}
\times i(S_{12}S_{34}-P_{12}P_{34})V_{56}^j
+\{c_1p_{12}^{-\,j}+c_2p_{56}^{-\,j}\}\times iV_{12}^kV_{34}^kV_{56}^j
\cr
&+e_1p_{12}^{+\,k}\epsilon^{j k}i(P_{12}S_{34}-S_{12}P_{34})V_{56}^j.\cr
{\cal S}^{(6,3)a}&=\{
  \ks_{1}p_{12}^{-\,j}p_{12}^-\cdot p_{12}^-
 +\ks_{2}p_{12}^{-\,j}p_{12}^-\cdot p_{34}^-
 +\ks_{3}p_{12}^{-\,j}p_{34}^+\cdot p_{34}^+
 +\ks_{4}p_{12}^{-\,j}p_{34}^+\cdot p_{56}^+
 +\ks_{5}p_{12}^{-\,j}p_{34}^-\cdot p_{34}^-\cr
&+\ks_{6}p_{12}^{-\,j}p_{34}^-\cdot p_{56}^-
 +\ks_{7}p_{34}^{+\,j}p_{12}^-\cdot p_{34}^+
 +\ks_{8}p_{34}^{+\,j}p_{12}^-\cdot p_{56}^+
 +\ks_{9}p_{34}^{+\,j}p_{34}^+\cdot p_{34}^-
 +\ks_{10}p_{34}^{+\,j}p_{34}^+\cdot p_{56}^-\cr
&+\ks_{11}p_{34}^{+\,j}p_{34}^-\cdot p_{56}^+
 +\ks_{12}p_{34}^{+\,j}p_{56}^+\cdot p_{56}^-
 +\ks_{13}p_{34}^{-\,j}p_{12}^-\cdot p_{12}^-
 +\ks_{14}p_{34}^{-\,j}p_{12}^-\cdot p_{34}^-
 +\ks_{15}p_{34}^{-\,j}p_{12}^-\cdot p_{56}^-\cr
&+\ks_{16}p_{34}^{-\,j}p_{34}^+\cdot p_{34}^+
 +\ks_{17}p_{34}^{-\,j}p_{34}^+\cdot p_{56}^+
 +\ks_{18}p_{34}^{-\,j}p_{34}^-\cdot p_{34}^-
 +\ks_{19}p_{34}^{-\,j}p_{34}^-\cdot p_{56}^-
 +\ks_{20}p_{34}^{-\,j}p_{56}^+\cdot p_{56}^+\cr
&+\ks_{21}p_{34}^{-\,j}p_{56}^-\cdot p_{56}^-\}\times
iV_{12}^j(S_{34}S_{56}-P_{34}P_{56}).\cr
{\cal S}^{(6,3)b}&=\{
  \kv_{1}p_{12}^{-\,j}p_{12}^-\cdot p_{12}^-
 +\kv_{2}p_{12}^{-\,j}p_{12}^-\cdot p_{34}^-
 +\kv_{3}p_{12}^{-\,j}p_{34}^+\cdot p_{34}^+
 +\kv_{4}p_{12}^{-\,j}p_{34}^+\cdot p_{56}^+
 +\kv_{5}p_{12}^{-\,j}p_{34}^-\cdot p_{34}^-\cr
&+\kv_{6}p_{12}^{-\,j}p_{34}^-\cdot p_{56}^-
 +\kv_{7}p_{34}^{+\,j}p_{12}^-\cdot p_{34}^+
 +\kv_{8}p_{34}^{+\,j}p_{12}^-\cdot p_{56}^+
 +\kv_{9}p_{34}^{+\,j}p_{34}^+\cdot p_{34}^-
 +\kv_{10}p_{34}^{+\,j}p_{34}^+\cdot p_{56}^-\cr
&+\kv_{11}p_{34}^{+\,j}p_{34}^-\cdot p_{56}^+
 +\kv_{12}p_{34}^{+\,j}p_{56}^+\cdot p_{56}^-
 +\kv_{13}p_{34}^{-\,j}p_{12}^-\cdot p_{12}^-
 +\kv_{14}p_{34}^{-\,j}p_{12}^-\cdot p_{34}^-
 +\kv_{15}p_{34}^{-\,j}p_{12}^-\cdot p_{56}^-\cr
&+\kv_{16}p_{34}^{-\,j}p_{34}^+\cdot p_{34}^+
 +\kv_{17}p_{34}^{-\,j}p_{34}^+\cdot p_{56}^+
 +\kv_{18}p_{34}^{-\,j}p_{34}^-\cdot p_{34}^-
 +\kv_{19}p_{34}^{-\,j}p_{34}^-\cdot p_{56}^-
 +\kv_{20}p_{34}^{-\,j}p_{56}^+\cdot p_{56}^+\cr
&+\kv_{21}p_{34}^{-\,j}p_{56}^-\cdot p_{56}^-\}\times
iV_{12}^j(S_{34}S_{56}-P_{34}P_{56}).\cr
{\cal S}^{(6,3)c}&=\{
  \jv_{1}p_{12}^{-\,j}p_{12}^{+\,k}p_{12}^{+\,l}
 +\jv_{2}p_{12}^{-\,j}p_{12}^{-\,k}p_{12}^{-\,l}
 +\jv_{3}p_{12}^{-\,j}p_{12}^{+\,k}p_{34}^{+\,l}
 +\jv_{4}p_{12}^{-\,j}p_{12}^{-\,k}p_{34}^{-\,l}
 +\jv_{5}p_{12}^{-\,j}p_{12}^{-\,k}p_{56}^{-\,l}\cr
&+\jv_{6}p_{34}^{-\,j}p_{12}^{+\,k}p_{12}^{+\,l}
 +\jv_{7}p_{34}^{+\,j}p_{12}^{+\,k}p_{12}^{-\,l}
 +\jv_{8}p_{56}^{+\,j}p_{12}^{+\,k}p_{12}^{-\,l}
 +\jv_{9}p_{34}^{-\,j}p_{12}^{-\,k}p_{12}^{-\,l}
 +\jv_{10}p_{12}^{-\,j}p_{34}^{-\,k}p_{56}^{-\,l}\cr
&+\jv_{11}p_{12}^{-\,j}p_{56}^{+\,k}p_{34}^{+\,l}
 +\jv_{12}p_{12}^{-\,j}p_{56}^{-\,k}p_{34}^{-\,l}
 +\jv_{13}p_{56}^{+\,j}p_{12}^{+\,k}p_{34}^{-\,l}
 +\jv_{14}p_{56}^{-\,j}p_{12}^{-\,k}p_{34}^{-\,l}\}\times
iV_{12}^jV_{34}^kV_{56}^l.\cr
{\cal S}^{(6,3)d}&=\{
  \ay_{1}p_{12}^{+\,j}p_{12}^+\cdot p_{12}^+
 +\ay_{2}p_{12}^{+\,j}p_{12}^+\cdot p_{34}^+
 +\ay_{3}p_{12}^{+\,j}p_{34}^+\cdot p_{34}^+
 +\ay_{4}p_{12}^{+\,j}p_{12}^-\cdot p_{12}^-
 +\ay_{5}p_{12}^{+\,j}p_{12}^-\cdot p_{34}^-\cr
&+\ay_{6}p_{12}^{+\,j}p_{12}^-\cdot p_{56}^-
 +\ay_{7}p_{12}^{+\,j}p_{34}^-\cdot p_{34}^-
 +\ay_{8}p_{12}^{+\,j}p_{34}^-\cdot p_{56}^-
 +\ay_{9}p_{12}^{+\,j}p_{56}^-\cdot p_{56}^-
 +\ay_{10}p_{12}^{-\,j}p_{12}^+\cdot p_{12}^-\cr
&+\ay_{11}p_{12}^{-\,j}p_{12}^+\cdot p_{34}^-
 +\ay_{12}p_{12}^{-\,j}p_{12}^+\cdot p_{56}^-
 +\ay_{13}p_{12}^{-\,j}p_{34}^+\cdot p_{12}^-
 +\ay_{14}p_{12}^{-\,j}p_{34}^+\cdot p_{34}^-
 +\ay_{15}p_{12}^{-\,j}p_{34}^+\cdot p_{56}^-\cr
&+\ay_{16}p_{56}^{-\,j}p_{12}^+\cdot p_{12}^-
 +\ay_{17}p_{56}^{-\,j}p_{12}^+\cdot p_{34}^-
 +\ay_{18}p_{56}^{-\,j}p_{12}^+\cdot p_{56}^- \}
 \times i\epsilon^{kj}(S_{12}P_{34}-P_{12}S_{34})V_{56}^k.\cr
{\cal S}^{(6,3)e}&=\{
  \by_{1}p_{12}^{+\,j}p_{12}^{+\,k}p_{34}^{+\,l}
 +\by_{2}p_{12}^{+\,j}p_{12}^{-\,k}p_{34}^{-\,l}
 +\by_{3}p_{12}^{+\,j}p_{12}^{-\,k}p_{56}^{-\,l}
 +\by_{4}p_{12}^{+\,j}p_{34}^{-\,k}p_{56}^{-\,l}
 +\by_{5}p_{12}^{-\,j}p_{12}^{+\,k}p_{12}^{-\,l}\cr
&+\by_{6}p_{12}^{-\,j}p_{12}^{+\,k}p_{34}^{-\,l}
 +\by_{7}p_{12}^{-\,j}p_{12}^{+\,k}p_{56}^{-\,l}
 +\by_{8}p_{12}^{-\,j}p_{34}^{+\,k}p_{12}^{-\,l}
 +\by_{9}p_{12}^{-\,j}p_{34}^{+\,k}p_{34}^{-\,l}
 +\by_{10}p_{12}^{-\,j}p_{34}^{+\,k}p_{56}^{-\,l}\cr
&+\by_{11}p_{56}^{-\,j}p_{12}^{+\,k}p_{12}^{-\,l}
 +\by_{12}p_{56}^{-\,j}p_{12}^{+\,k}p_{34}^{-\,l}
 +\by_{13}p_{56}^{-\,j}p_{12}^{+\,k}p_{56}^{-\,l}
\}\times i\epsilon^{kl}(S_{12}P_{34}-P_{12}S_{34})V_{56}^j
.&\sss \cr}$$

The conventions to get rid of non-independent operators are as follows.
In ${\cal S}^{(4,2)}$ we consider terms containing $p_{34}^{+\,k}$
multiplied by neither $(S_{12}S_{34} -P_{12}P_{34})$ nor $V_{12}^jV_{34}^j$,
because, for instance, $p_{12}^+\cdot p_{34}^+V_{12}^jV_{34}^j=-p_{12}^{+\,2}
V_{12}^jV_{34}^j$.  We do not take into account operators containing
$(S_{12}S_{34}-P_{12}P_{34})V_{56}^jp_{56}^{+\,k}$, $V_{12}^kV_{34}^kV_{56}^j
p_{56}^{+\,l}$ or $(S_{12}P_{34}-P_{12}S_{34})V_{56}^jp_{56}^{+\,k}$ either.
Finally, for terms like $V_{12}^jV_{34}^k$ we integrate by parts if they
are multiplied by $p_{12}^{+\,j}$ or $p_{34}^{+\,k}$, e.g., $V_{12}^j
V_{34}^kp_{12}^{+\,j}p_{34}^{+\,k}=-V_{12}^jV_{34}^kp_{34}^{+\,j}p_{34}^{+\,k}
=V_{12}^jV_{34}^kp_{34}^jp_{12}^{+\,k}$, and similarly, for $V_{12}^j
V_{34}^kV_{56}^l$ multiplied by $p_{12}^{+\,j}$, $p_{34}^{+\,k}$ or
$p_{56}^{+\,l}$.

We do not include ${\cal S}^{(2,3)}=iV_{12}^j(p_1-p_2)^j(p_1-p_2)^2$
as we have discussed in
Sect.~3.  Without it, we have a basis of 106 independent operators.

\appendix{C}{$\beta$ functions}
In this appendix we present the complete set of $\beta$-functions.
The scheme dependent parameters $\alpha$, $\beta$, $\gamma$ and
$\delta$ have been defined in Eq.~\spar .

\eqnn\betaca
$$\eqalignno{\eta&
= 4 \alpha [-m_1 + m_2 + m_3 + s_1 + s_2 + s_3 + s_4 + t - 2 N ( r_2 +
     2 s_3 + s_4 )],\cr
}$$
$$\eqalignno{
{\dot g}_1&= 2 g_1 \eta + 2 \alpha ( a_1 + a_2 + e_1 - a_2 N) + 2 \beta
(-\ay_1
- \ay_4 - \ay_6 - \ay_9 - \ay_{10} - \ay_{12} - \ay_{16} - \ay_{18}
\cr &+ 2 \ks_1
+ \ks_2 + \ks_3 + \ks_5 + \ks_7 + \ks_9 + \ks_{13} + \ks_{14}
+ \ks_{16} + \ks_{18} - 8 \ks_1 N),\cr
{\dot g}_2 &= 2 g_2 \eta + 2 \alpha ( a_1 + \cc_1 + \cc_2 + e_1 - \cc_1 N -
\cc_2 N) + \beta [ 2 ( - \ay_1
 + \ay_2 - \ay_3- \ay_4 - \ay_5 - \ay_7
- \ay_{10} - \ay_{11} + \ay_{13} +
 \cr &\ay_{14} )
+ 2 \jv_1 + 2
\jv_2 - \jv_3 + \jv_4 + \jv_5 + \jv_6 - \jv_7 - \jv_8 + \jv_9
+ 2 ( \ks_9 +
\ks_{10} - \ks_{11} - \ks_{12} + \ks_{16} - \ks_{17} +
\ks_{18} \cr & + \ks_{19} + \ks_{20} + \ks_{21}
) + 2 ( 2 \kv_1 +
\kv_{13}
 + \kv_{14} + \kv_{16} + \kv_{18} + \kv_2 + \kv_3 +
\kv_5 + \kv_7 + \kv_9 ) - 8 N ( \jv_2 + 2 \kv_1 + \kv_{18} )],\cr
{\dot m}_1 &= (-2 + 2 \eta) m_1 + {\alpha\over 2} [-2 \ay_1
+ 5 \ay_2 - 6 \ay_3 +\ay_6 + 2 \ay_7 - 4 \ay_9
+ \ay_{10} - 3 \ay_{13} + \ay_{15} + \ay_{16} + \ay_{17} - 5 \ay_{18}
\cr &+ \by_1
+ \by_3 + 2 \by_4 + \by_5 - \by_8
- \by_{10} - \by_{11} - \by_{12} - \by_{13}
+ 4 \ks_1
- \ks_2 + 6 \ks_3 - 6 \ks_4 + 4 \ks_7 - 5 \ks_8 - \ks_9
\cr & + 2 \ks_{10} + \ks_{11} - \ks_{14} - \ks_{17} + 2 \ks_{18} + 4 \ks_{20}
+ 4 N ( - 2 \ks_3 + 2 \ks_4 - \ks_7
 + \ks_8 )],\cr
{\dot m}_2 &= (-2 + 2 \eta) m_2 + {\alpha\over 2} [- \ay_5 - \ay_8  +
\ay_{11} + \ay_{17} - \by_2 + \by_4 + \by_6 - 2 \by_9 - 2 \by_{10} + \by_{12}
\cr &
+ 3 \ks_2 + 6 \ks_6 - 3 \ks_{10} + 2 \ks_{13} + 5 \ks_{15} + 3 \ks_{19} - 2
\ks_{20} + 2 \ks_{21} - 4 N ( 2 \ks_6 + \ks_{15} )],\cr
{\dot m}_3 &= (-2 + 2 \eta) m_3 +{\alpha\over 2} [-2 \ay_1 - 4 \ay_7
+ \ay_{10} + \ay_{12} + 2 \ay_{13} + \ay_{16} + \ay_{18} + \by_5 + \by_7
 - 2 \by_8 + \by_{11} + \by_{13} \cr
 &+ 4 \ks_1 + 2 \ks_2 + 6 \ks_5 - \ks_7 - \ks_9 + 2 \ks_{13} + 4 \ks_{14} +
2 \ks_{18} + 2 \ks_{19} + 4 \ks_{21}
 - 4 N ( 2 \ks_5 + \ks_{14} )],\cr
{\dot r}_1 &= (-2 + 2 \eta)r_1 + {\alpha\over 4} [ 2 ( - 3 \ay_1
- \ay_2 + \ay_3 + \ay_4 + \ay_5 - 3 \ay_7
+ \ay_{10} + \ay_{11} + 3 \ay_{13} - \ay_{14}
) \cr & + 2 \jv_1 - 2 \jv_2 - 3 \jv_3
- \jv_4 + \jv_5 - 5 \jv_6 - 3 \jv_7 + 3 \jv_8 + \jv_9
+ 4 \jv_{11} + 4 \jv_{13} \cr & + 2 ( - \ks_9 + \ks_{10} - \ks_{11}
 + \ks_{12} + \ks_{16} + \ks_{17} + \ks_{18}
- \ks_{19} + \ks_{20} + \ks_{21}
) + 2 ( 4 \kv_1 - \kv_2
+ 6 \kv_3 - 6 \kv_4 \cr & + 4 \kv_7 - 5 \kv_8 - \kv_9
+ 2 \kv_{10} + \kv_{11} - \kv_{14} - \kv_{17} + 2 \kv_{18}
+ 4 \kv_{20} ) + 8 N ( - 2 \kv_3 + 2 \kv_4 - \kv_7 + \kv_8
- \kv_{20} )],\cr
{\dot r}_2 &= (-2 + 2 \eta) r_2 + {\alpha\over 2} [- \jv_4 + \jv_6 - \jv_9
- 2 \jv_{12} - \jv_{13} - 3 \jv_{14} + 2 ( - \ks_7
+ \ks_8 + \ks_{14} + \ks_{15} ) \cr
 & + 3 \kv_2 + 6 \kv_6 - 3 \kv_{10} + 2 \kv_{13} + 5 \kv_{15} + 3 \kv_{19}
- 2 \kv_{20} + 2 \kv_{21} - 8 N ( \kv_6 + \kv_{15} )],\cr
{\dot r}_3 &= (-2 + 2 \eta) r_3 + {\alpha\over 4} [ 2 ( - 3 \ay_1
+ 3 \ay_2 - 3 \ay_3 + \ay_4 + \ay_5 + \ay_7 - 4 \ay_9
+ \ay_{10} + \ay_{11} - \ay_{13} - \ay_{14} - 4 \ay_{18}
) + 2 \jv_1 - 2 \jv_2 \cr & - \jv_3 -
\jv_4 - \jv_5 + \jv_6 + \jv_7 - \jv_8 - 5 \jv_9 + 2 ( - \ks_9
- \ks_{10} + \ks_{11} + \ks_{12}
+ 4 \ks_{13} + \ks_{16} - \ks_{17}
+ \ks_{18} + \ks_{19} + \ks_{20} \cr & + \ks_{21}
) + 2 ( 4 \kv_1 + 2 \kv_2
+ 6 \kv_5 - \kv_7 - \kv_9
+ 2 \kv_{13} + 4 \kv_{14} + 2 \kv_{18} + 2 \kv_{19} + 4 \kv_{21}
) - 8 N ( 2 \kv_5 + \kv_{13} + \kv_{14} + \kv_{21} )],\cr
{\dot s}_1 &= (-2 + 2 \eta) s_1 + {\alpha\over 2} [ 2 ( - \ay_1
+ \ay_3 + \ay_4 - \ay_7 + \ay_{11}
+ \ay_{13} ) - 2 ( \by_1 + \by_2 + \by_5
+ \by_9 )
\cr & + 2 \jv_1 - 6 \jv_2 - \jv_3 + \jv_4
+ \jv_5 - 7 \jv_6 - \jv_7 - \jv_8 + \jv_9 + 8 \jv_{11} + 8 \jv_{13}
\cr & + 2 ( - \ks_{10} - \ks_{11} +
\ks_{16} - \ks_{18} - \ks_{20} + \ks_{21} ) + 4 N ( \jv_6 - 2 \jv_{11} -
\jv_{13} + \kv_{10} )],\cr
{\dot s}_2 &= (-2 + 2 \eta) s_2 +{\alpha\over 2}[2 ( \ay_1 -
\ay_2 + \ay_3 - \ay_4 - \ay_5 - \ay_7 + 2 \ay_{18}
) + 2 ( \by_5 + \by_6 - \by_8 - \by_9 + 2 \by_{13}
) \cr & - 2 \jv_1 + 6 \jv_2 + \jv_3 + 7 \jv_4 + 7
\jv_5 - \jv_6 + \jv_7 + \jv_8 + 7 \jv_9
\cr & + 2 ( 2 \ks_2 - \ks_{16} + \ks_{17} +
\ks_{18} + \ks_{19} - \ks_{20} + \ks_{21} ) - 4 N ( \jv_4
+ 2 \jv_5 + \jv_9 + \kv_2 + \kv_{19} )],\cr
{\dot s}_3 &= (-2 + 2 \eta) s_3 + {\alpha\over 2} [ 2 ( - \ay_6 - \ay_8
+ \ay_{12} - \ay_{15}
) + 2 ( - \by_3 - \by_4 + \by_7
- \by_{10} + 2 \by_{11} + 2 \by_{12} )
\cr & + \jv_4 + 4 \jv_5 - \jv_6 + \jv_9
 + 12 \jv_{10} + 4 \jv_{11} + 6 \jv_{12} + \jv_{13} + 3 \jv_{14}
\cr & + 4 ( - \ks_3 + \ks_4 + \ks_5 + \ks_6
 ) - \kv_{10} - 2 \kv_{13} - \kv_{15} + \kv_{19}
+ \kv_2 + 2 \kv_6 + 2 \kv_{20} - 2 \kv_{21} - 8 N ( 3 \jv_{10} + \jv_{12}
+ \kv_6 )],\cr
{\dot s}_4 &= (-2 + 2 \eta) s_4 + {\alpha\over 2} [3 ( \jv_4 - \jv_6 + \jv_9
+ 2 \jv_{12} + \jv_{13} + 3 \jv_{14} ) + 2 ( - \ks_7 + \ks_8 + \ks_{14}
+ \ks_{15} ) \cr &- \kv_2 - 2 \kv_6 + \kv_{10} + 2 \kv_{13} + \kv_{15} -
\kv_{19} - 2 \kv_{20} + 2 \kv_{21} - 4 N ( 2 \jv_{12} + 3
\jv_{14}
 + \kv_{15} )],\cr
{\dot t} &= (-2 + 2 \eta) t + {\alpha\over 2} (
\ay_2 + 2 \ay_3 - 3 \ay_5 - 3 \ay_6 - 2 \ay_7 - 3 \ay_8 - 4 \ay_9
+ 3 \ay_{11} + 3 \ay_{12} - \ay_{13} - 3 \ay_{15} + 2 \ay_{18}
\cr & + 3 ( - \by_1
- \by_2 - \by_3 - \by_4 + \by_6 + \by_7 - \by_8 - 2 \by_9
- \by_{10} + 2 \by_{11} + 2 \by_{12} + 2 \by_{13}
) \cr & + 2 \ks_2 - 2 \ks_3 + 2 \ks_4 + 2 \ks_5 + 2 \ks_6 + \ks_7 - \ks_8
+ \ks_{10} - 3 \ks_{11} - 4 \ks_{13} - \ks_{14} -
\ks_{15} + 3 \ks_{17} - \ks_{19} - 2 \ks_{20} + 2 \ks_{21}
\cr &+ 4 N ( \ay_6 + \ay_8 - \ay_{12} + \ay_{15}
+ \by_3 + \by_4 - \by_7
+ \by_{10} - 2 \by_{11} - 2 \by_{12} )
),\cr
}$$
$$\eqalignno{
{\dot a}_1&= (-2 + 3 \eta) a_1 + 4 g_1 g_2 \gamma,\ \
{\dot a}_2= (-2 + 3 \eta) a_2 + 2 g_1^2 \gamma,\ \
{\dot \cc}_{1}= (-2 + 3 \eta) \cc_1 + 4 g_2^2 \gamma,\cr
{\dot \cc}_{2}& = (-2 + 3 \eta) \cc_2 - 2 g_2^2 \gamma,\ \
{\dot e}= (-2 + 3 \eta) e - 4 g_1^2 \gamma - 4 g_1 g_2 \gamma ,\cr
}$$
$$\eqalignno{
{\dot \ks_1} &=(-4 + 3 \eta) \ks_1 + \delta g_1^2/2 + 2 g_1 \gamma
m_3,\ \
{\dot \ks_2} = (-4 + 3 \eta) \ks_2 + 2 \gamma g_1 (2 m_2 + s_2),\cr
{\dot \ks_3} &= (-4 + 3 \eta) \ks_3 + \delta g_1^2 + 2 \gamma g_1 (2 m_1 +
m_3 + 4 s_3 + t), \ \
{\dot \ks_4} = (-4 + 3 \eta) \ks_4 -  \delta g_1^2 + 2 \gamma g_1 (2 s_3 -
t),\cr
{\dot \ks_5} &= (-4 + 3 \eta) \ks_5 + 2 \gamma g_1 (m_3 + 2 s_3),\ \
{\dot \ks_6} = (-4 + 3 \eta) \ks_6 \cr
{\dot \ks_7} &= (-4 + 3 \eta) \ks_7 + 2 \delta g_1^2 + 2 \gamma g_1 (2 m_3
+ 4 r_2 + 4 s_4 - t),\cr
{\dot \ks_8} &= (-4 + 3 \eta) \ks_8 -2 \delta g_1^2 + 2 \gamma g_1 (-2 m_3
+ 2 r_2 + 2 s_4 + t), \cr
{\dot \ks_9} &= (-4 + 3 \eta) \ks_9 + 2 \delta g_1 g_2 + 2 \gamma (
- g_1 ( 2 s_1 + t ) + 2 g_2 ( m_2 + m_3 - t )) ,\cr
{\dot \ks_{10}} &= (-4 + 3 \eta) \ks_{10} + 8 \delta g_1 g_2 + 2 \gamma (
g_1 ( 2 m_2 + 2 r_3 - 2 s_1 + 3 s_2 ) + 2 g_2 ( m_2 + 2 m_3
- t ) ), \cr
{\dot \ks_{11}} &=(-4 + 3 \eta) \ks_{11} + 4 \delta g_1 g_2 + 2 \gamma (
g_1 ( - 2 m_2 + 2 r_3 - 2 s_1 + t ) + 2 g_2 ( 2 m_2 + m_3 ) )
,\cr
{\dot \ks_{12}} &=(-4 + 3 \eta) \ks_{12} + 4 \delta g_1 g_2 + 2 \gamma (
g_1 ( - 2 s_1 + s_2 ) + 2 g_2 ( m_2 + 2 m_3 - t) ),\ \
{\dot \ks_{13}} = (-4 + 3 \eta) \ks_{13} + 2 \gamma g_1 r_3, \cr
{\dot \ks_{14}} &= (-4 + 3 \eta) \ks_{14} + 4 \gamma g_1 ( r_2 + s_4),\ \
{\dot \ks_{15}} = (-4 + 3 \eta) \ks_{15},\cr
{\dot \ks_{16}} &= (-4 + 3 \eta) \ks_{16} +\delta  g_1 g_2 + 2 \gamma (
g_1 ( 2 r_1 + t ) + g_2 ( m_3 + 2 t) ) ,\cr
{\dot \ks_{17}} &= (-4 + 3 \eta) \ks_{17} + 4 \delta g_1 g_2 + 2 \gamma (
g_1 ( 4 r_1 + s_2 - t ) + 4 g_2 ( m_3 + t) ),\cr
{\dot \ks_{18}} &= (-4 + 3 \eta) \ks_{18} + \delta g_1 g_2 + 2 \gamma (
g_1 ( r_3 + s_2 ) + g_2 m_3 ),\cr
{\dot \ks_{19}} & = (-4 + 3 \eta) \ks_{19} + 4 \gamma g_2 m_2,\ \
{\dot \ks_{20}} =(-4 + 3 \eta) \ks_{20} + 4 \delta g_1 g_2 + \gamma (
g_1 ( 2 r_1 + r_3 + 2 s_2 ) + g_2 ( 2 m_1 + m_3 + 2 t) ),\cr
{\dot \ks_{21}} &= (-4 + 3 \eta) \ks_{21} + 2 \gamma g_2 m_3,\ \
{\dot \kv_1 } = (-4 + 3 \eta) \kv_1 - \delta g_2^2/2 - 2 \gamma g_2 r_3,
\ \
{\dot \kv_2 } = (-4 + 3 \eta) \kv_2 + 2 \gamma g_2 (-2 r_2 + s_2)
,\cr
{\dot \kv_3 } &= (-4 + 3 \eta) \kv_3 - \delta g_2^2 + 2 \gamma g_2 (-2 r_1
- r_3 + 4 s_3), \ \
{\dot \kv_4 } = (-4 + 3 \eta) \kv_4 + \delta g_2^2 + 4 \gamma g_2 s_3
,\cr
{\dot \kv_5 } &= (-4 + 3 \eta) \kv_5 + 2 \gamma g_2 (- r_3 + 2 s_3)
,\ \
{\dot \kv_6 } = (-4 + 3 \eta) \kv_6 ,\ \
{\dot \kv_7 } = (-4 + 3 \eta) \kv_7 -2 \delta g_2^2 + 4 \gamma g_2 (2 r_2
- r_3 + 2 s_4), \cr
{\dot \kv_8 } &= (-4 + 3 \eta) \kv_8 + 2 \delta  g_2^2+ 4 \gamma g_2 (
r_2 + r_3 + s_4), \ \
{\dot \kv_9 } = (-4 + 3 \eta) \kv_9 + 2 \delta g_2^2 + 4 \gamma g_2 ( 2
r_2 + r_3 - s_1), \cr
{\dot \kv_{10} } &= (-4 + 3 \eta) \kv_{10} + 8 \delta g_2^2
+ 2 \gamma g_2 ( 8 r_3 - 2 s_1 + 3 s_2), \ \
{\dot \kv_{11} } = (-4 + 3 \eta) \kv_{11} + 4  \delta g_2^2 + 4 \gamma g_2 (3
r_2 + 2 r_3 - s_1) ,\cr
{\dot \kv_{12} } &= (-4 + 3 \eta) \kv_{12} + 4 \delta g_2^2 + 2 \gamma g_2 (2
r_2 + 4 r_3 - 2 s_1 + s_2) ,\ \
{\dot \kv_{13} } = (-4 + 3 \eta) \kv_{13} + 2 \gamma g_2 r_3 ,\cr
{\dot \kv_{14} } &= (-4 + 3 \eta) \kv_{14} + 4 \gamma g_2 (r_2 +
s_4), \ \
{\dot \kv_{15} } = (-4 + 3 \eta) \kv_{15},\ \
{\dot \kv_{16} } = (-4 + 3 \eta) \kv_{16} + \delta g_2^2 + 2 \gamma g_2 (2
r_1 + r_3) ,\cr
{\dot \kv_{17} } &= (-4 + 3 \eta) \kv_{17} + \delta 4 g_2^2 + 2 \gamma g_2 (4
r_1 + 2 r_3 + s_2), \ \
{\dot \kv_{18} } = (-4 + 3 \eta) \kv_{18} + \delta g_2^2 + 2 \gamma g_2 (2
r_3 + s_2),\cr
{\dot \kv_{19} } &= (-4 + 3 \eta) \kv_{19} + 4 \gamma g_2 r_2,\ \
{\dot \kv_{20} } = (-4 + 3 \eta) \kv_{20} + 4 g_2^2 \delta + 4 \gamma g_2 (2
r_1 + r_3 + s_2), \cr
{\dot \kv_{21} } &= (-4 + 3 \eta) \kv_{21} + 2 \gamma g_2 r_3,\ \
 {\dot \jv_1} = (-4 + 3 \eta) \jv_1 - 4 \gamma g_2 s_3,\ \
 {\dot \jv_2} = (-4 + 3 \eta) \jv_2,\cr
 {\dot \jv_3} &= (-4 + 3 \eta) \jv_3 - 4 \gamma g_2 (s_1 + 2  s_3), \ \
 {\dot \jv_4} = (-4 + 3 \eta) \jv_4 + 2 \gamma g_2 (s_2 - 2 s_4), \ \
 {\dot \jv_5} = (-4 + 3 \eta) \jv_5 -2 \gamma  g_2 s_2 ,\cr
 {\dot \jv_6} &= (-4 + 3 \eta) \jv_6 - 2 \gamma g_2 (s_2 + 2 s_4), \ \
 {\dot \jv_7} = (-4 + 3 \eta) \jv_7 + 4 \gamma g_2 (s_1 - 2 s_4) ,\cr
 {\dot \jv_8} &= (-4 + 3 \eta) \jv_8 + 2 \gamma g_2 (- s_2 + 2 s_4), \ \
 {\dot \jv_9} = (-4 + 3 \eta) \jv_9 + 4  \gamma  g_2  s_4, \ \
 {\dot \jv_{10}} = (-4 + 3 \eta) \jv_{10},\cr
 {\dot \jv_{11}} &= (-4 + 3 \eta) \jv_{11}
- 2 \gamma g_2 (2 s_1 + s_2),\ \
 {\dot \jv_{12}} = (-4 + 3 \eta) \jv_{12},\ \
 {\dot \jv_{13}} = (-4 + 3 \eta) \jv_{13}
+ 4 \gamma g_2 (s_1 - s_4), \cr
 {\dot \jv_{14}} &= (-4 + 3 \eta) \jv_{14},\ \
{\dot \ay_1} = (-4 + 3 \eta) \ay_1 + \delta g_1 ( g_1 + 7 g_2) +
2 \gamma (g_1 ( 2 m_1 + m_3 + 2 r_1 + 2 r_3 ) + g_2 ( 4 m_1 + m_3 ) ),\cr
{\dot \ay_2} &= (-4 + 3 \eta) \ay_2 + 2 \delta g_1 (-g_1 + 2 g_2) +
4 \gamma ( 2 g_1 r_1 + g_2 m_3 ),\cr
{\dot \ay_3} &= (-4 + 3 \eta) \ay_3 + \delta g_1 ( g_1 - 2 g_2) +
2 \gamma ( g_1 ( 2 m_1 + m_3 + 2 r_1 - r_3 ) + g_2 ( - 2 m_1 + m_3 ) ),\cr
{\dot \ay_4} &= (-4 + 3 \eta) \ay_4 - \delta g_1 g_2 + 2 \gamma (g_1 ( m_3
- r_3 ) + g_2 m_3 ), \ \
{\dot \ay_5} = (-4 + 3 \eta) \ay_5 + 4 \gamma g_2 m_2,\cr
{\dot \ay_6} &= (-4 + 3 \eta) \ay_6 + 4 \gamma g_1 (m_2 - r_2),\ \
{\dot \ay_7} = (-4 + 3 \eta) \ay_7 + 2 \delta g_1 g_2 + 2 \gamma (g_1
( m_3 + 2 r_3 ) + g_2 m_3 ),\cr
{\dot \ay_8} &= (-4 + 3 \eta) \ay_8 + 4 \gamma g_1 ( m_2 + 2 r_2),\ \
{\dot \ay_9} = (-4 + 3 \eta) \ay_9 + \delta g_1^2 + 2 \gamma g_1 (2 m_3 +
r_3),\cr
{\dot \ay_{10}} &= (-4 + 3 \eta) \ay_{10} -2 \delta g_1 g_2 - 4 \gamma
g_2 m_3,\ \
{\dot \ay_{11}} = (-4 + 3 \eta) \ay_{11} - 4 \gamma g_2 m_2,\ \
{\dot \ay_{12}} = (-4 + 3 \eta) \ay_{12},\cr
{\dot \ay_{13}} &= (-4 + 3 \eta) \ay_{13} -4 \delta g_1 g_2 - 4 \gamma (
g_1 r_3 + g_2 m_3 ),\ \
{\dot \ay_{14}} = (-4 + 3 \eta) \ay_{14} - 4 \gamma g_2 m_2,\cr
{\dot \ay_{15}} &= (-4 + 3 \eta) \ay_{15} - 4 \gamma g_1 r_2,\ \
{\dot \ay_{16}} = (-4 + 3 \eta) \ay_{16} ,\ \
{\dot \ay_{17}} = (-4 + 3 \eta) \ay_{17} + 4 \gamma g_1 m_2,\cr
{\dot \ay_{18}} &= (-4 + 3 \eta) \ay_{18} + 2 \delta g_1^2 + 4 \gamma
g_1 m_3,\ \
{\dot \by_1} = (-4 + 3 \eta) \by_1 + 2 \gamma (g_1 ( 4 s_1 + s_2 ) + 3
g_2 t),\ \
{\dot \by_2} = (-4 + 3 \eta) \by_2,\cr
{\dot \by_3} &= (-4 + 3 \eta) \by_3,\ \
{\dot \by_4} = (-4 + 3 \eta) \by_4 - 4 \gamma g_1 s_4,\ \
{\dot \by_5} = (-4 + 3 \eta) \by_5 + 2 \gamma g_1 s_2,\ \
{\dot \by_6} = (-4 + 3 \eta) \by_6,\cr
{\dot \by_7} &= (-4 + 3 \eta) \by_7 + 4 \gamma g_1 s_4,\ \
{\dot \by_8} = (-4 + 3 \eta) \by_8 + 2 \gamma (g_1 s_2 - g_2 t),\ \
{\dot \by_9} = (-4 + 3 \eta) \by_9 - 2 \gamma g_2 t,\cr
{\dot \by_{10}} &= (-4 + 3 \eta) \by_{10} + 8  \gamma g_1 s_4,\ \
{\dot \by_{11}} = (-4 + 3 \eta) \by_{11} + 2 \gamma g_1 ( 2 s_3 +
t),\ \
{\dot \by_{12}} = (-4 + 3 \eta) \by_{12} - 4 \gamma g_1 s_3,\cr
{\dot \by_{13}} &= (-4 + 3 \eta) \by_{13} + 2 \gamma g_1 (- s_2 +
t).
&\betaca \cr}$$

The only
manifestation of the cut-off function comes through the constants $\alpha$,
$\beta$, $\gamma$ and $\delta$.  However, as it happens in the bosonic case
\BHLM\ we can reduce the number of independent parameters from four to two.
In fact, by performing the following rescalings
\eqn\rescaling{
g^{(4,0)} \to {1\over \alpha \gamma} g^{(4,0)},\ \
g^{(4,2)} \to {\delta \over \alpha \gamma^2} g^{(4,2)},\ \
g^{(6,1)} \to {1\over \alpha^2 \gamma} g^{(6,1)},\ \
g^{(6,3)} \to {\delta \over \alpha^2 \gamma^2} g^{(6,3)},
}
where by $g^{(m, n)}$ we denote the coupling
constants corresponding to the operators with $m$ fermions
and $n$ derivatives,
it can be shown that the $\beta$-functions depend on the scheme
only through the combinations
$z={\delta \over \gamma^2}$ and
$w= {\beta\delta \over \alpha \gamma}$.
Moreover, $z$ enters the equations only as a global factor of the
anomalous dimension.

\footatend\vfill\supereject\immediate\closeout\rfile\writestoppt
\baselineskip=14pt\centerline{{\bf References}}\bigskip{\frenchspacing%
\parindent=20pt\escapechar=` \input refs.tmp\vfill\eject}\nonfrenchspacing
\vfill\eject\immediate\closeout\ffile{\parindent40pt
\baselineskip14pt\centerline{{\bf Figure Captions}}\nobreak\medskip
\escapechar=` \input figs.tmp\vfill\eject}
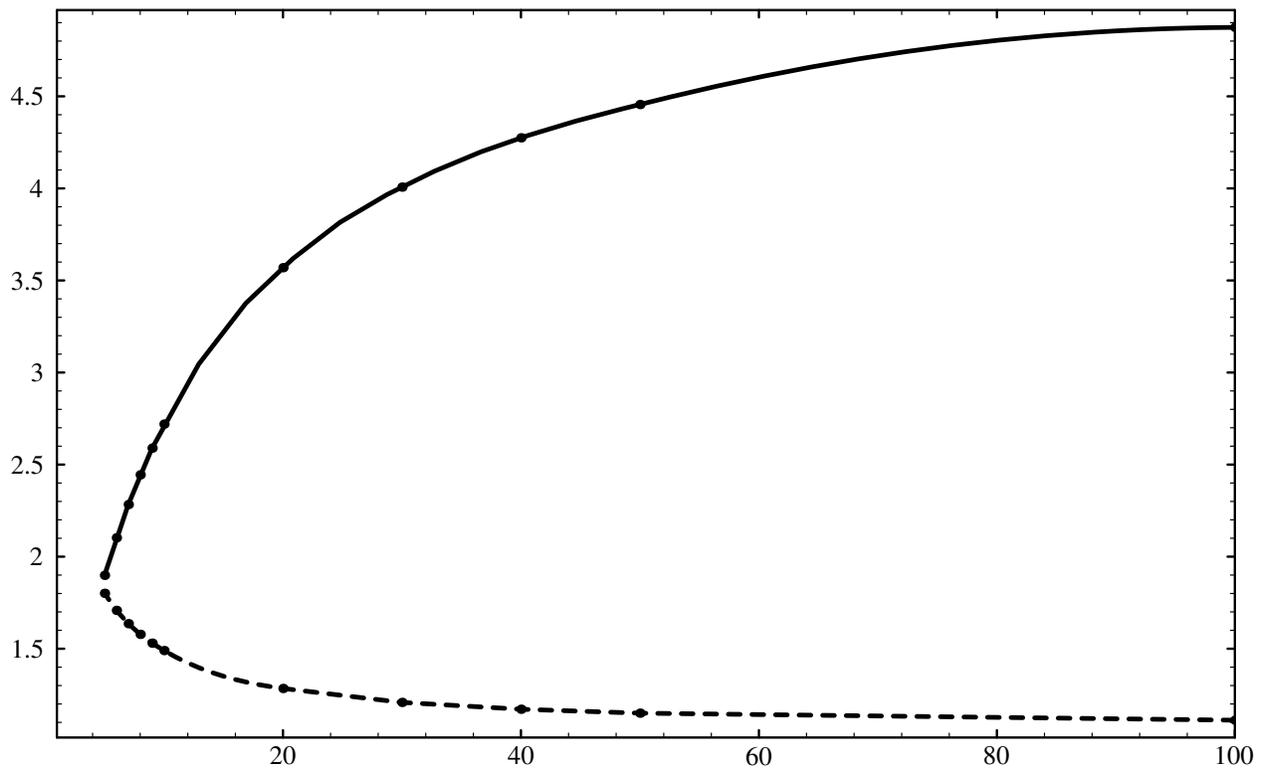
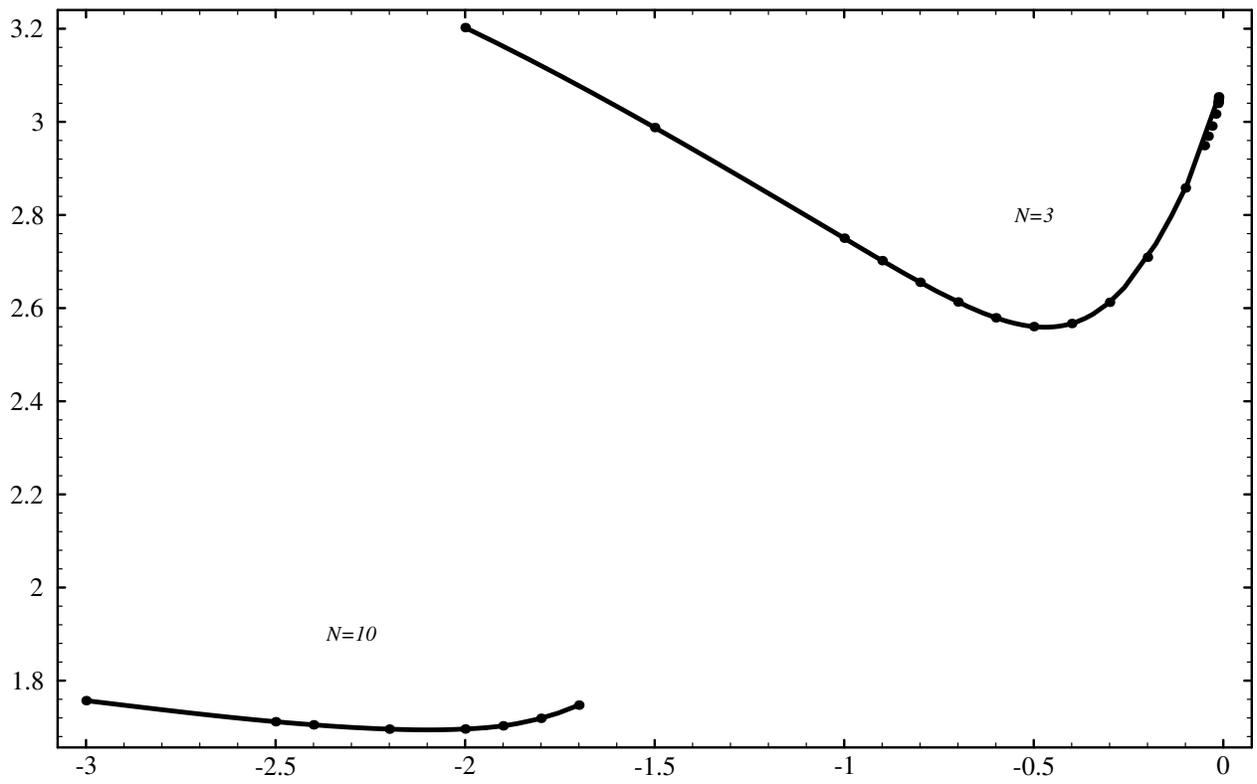
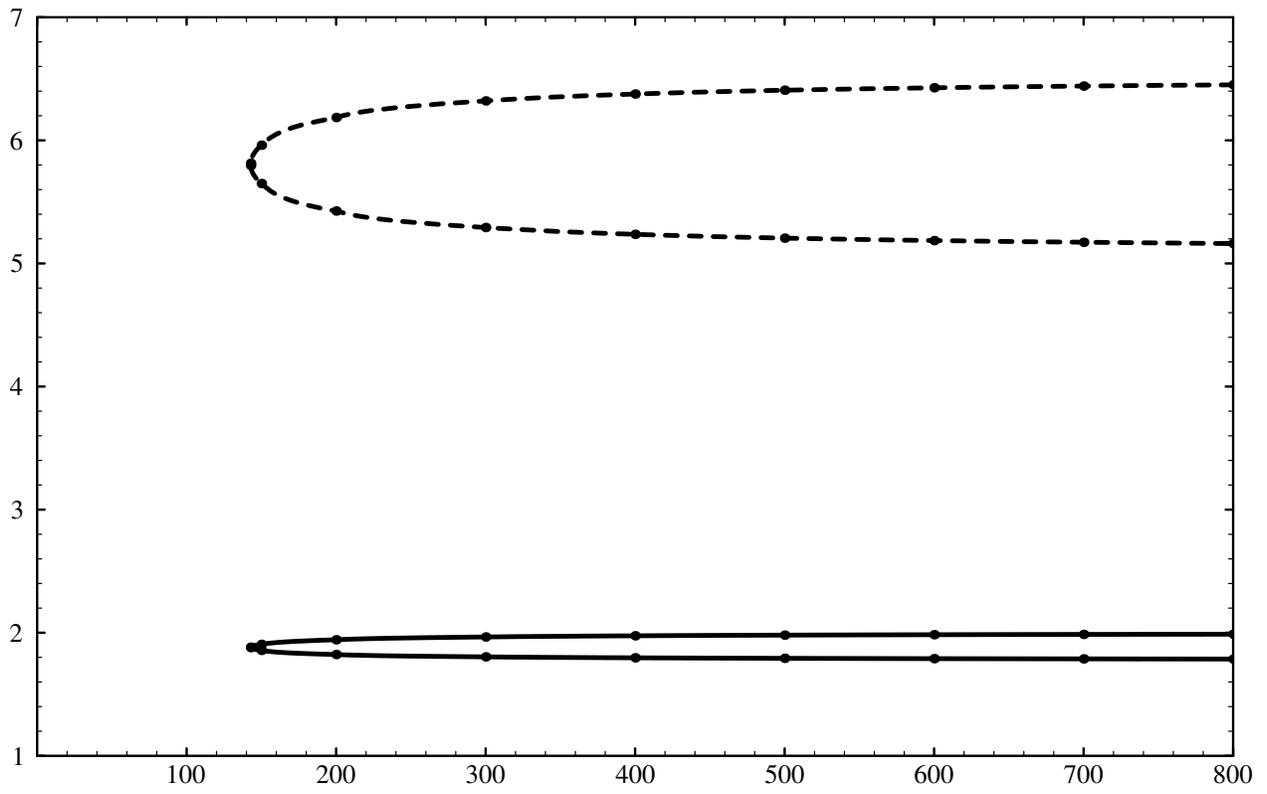
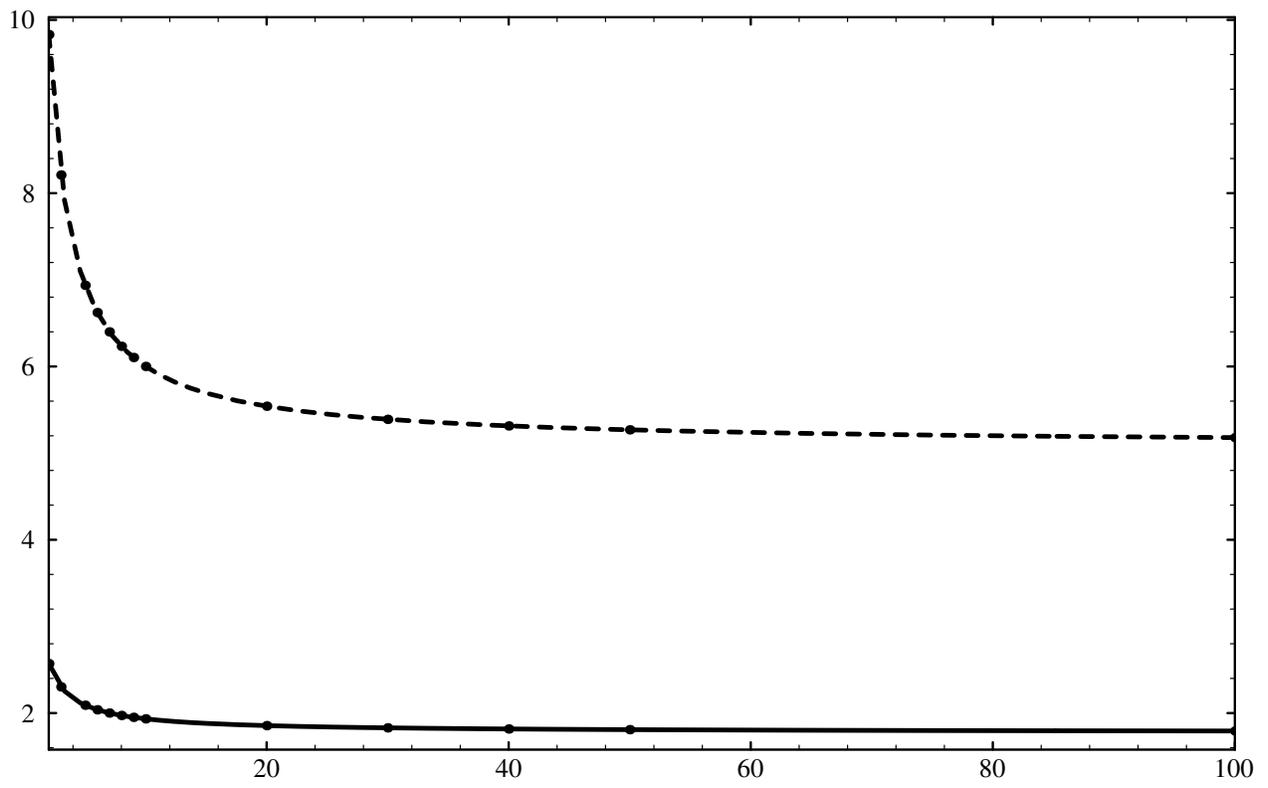
\bye